\documentclass[moor,nonblindrev]{informs1}
\RequirePackage[OT1]{fontenc}
\usepackage[inline]{enumitem}
\usepackage{bm,dsfont,nicefrac,upgreek}
\usepackage{mathtools}
\usepackage{subfig,wrapfig,epstopdf}
\usepackage[margin=10pt,font=small,labelfont=bf]{caption}
\usepackage[mathscr]{eucal}
\usepackage[normalem]{ulem}

\usepackage{makecell,longtable}
\usepackage{algorithm,algpseudocode}
\usepackage[sort]{natbib}
 \NatBibNumeric
 \def\bibfont{\small}%
 \bibpunct[, ]{[}{]}{,}{n}{}{,}%
\usepackage[colorlinks=true,breaklinks=true,bookmarks=true,urlcolor=blue,
     citecolor=blue,linkcolor=blue,bookmarksopen=false,draft=false]{hyperref}

\def\EMAIL#1{\href{mailto:#1}{#1}}

\TheoremsNumberedBySection
\EquationsNumberedThrough

\theoremstyle{EX}
\newtheorem{observation}{Observation}

\newcounter{const-no}
\newcommand{\const}[1]{\refstepcounter{const-no}\label{#1}}
\newcounter{event-no}
\newcommand{\event}[1]{\refstepcounter{event-no}\label{#1}}

\DeclarePairedDelimiterX{\card}[1]{\lvert}{\rvert}{#1}
\DeclarePairedDelimiterX{\norm}[1]{\lVert}{\rVert}{#1}
\DeclareMathOperator{\diag}{diag}

\newcommand*\diff{\mathop{}\!\mathrm{d}}
\newcommand{\ind}[1]{\mathds{1}\bigl\lbrace #1 \bigr\rbrace}
\newcommand\givenbase[1][]{\:#1\lvert\:}
\let\given\givenbase

\DeclarePairedDelimiterX\Basics[1](){\let\given\sgiven #1}

\newcommand{\EE}{{\mathbb{E}}}       
\newcommand{\PP}{{\mathbb{P}}}       
\newcommand{\RR}{\mathbb{R}}       
\newcommand{\NN}{\mathbb{N}}       
\newcommand{\ZZ}{\mathbb{Z}}       

\newcommand{\E}{\mathrm{e}}     
\newcommand{\Ind}{\mathds{1}}   
\newcommand{\df}{\coloneqq}     

\newcommand{\Lyap}{\mathscr{V}} 
\newcommand{\sA}{\mathscr{A}} 
\newcommand{\sB}{\mathscr{B}} 
\newcommand{\sP}{\mathscr{P}} 

\newcommand{\sE}{{\mathscr{E}}} 
\newcommand{\sG}{{\mathscr{G}}}  
\newcommand{\sI}{\mathscr{I}} 
\newcommand{\sJ}{\mathscr{J}}

\newcommand{\abs}[1]{\lvert#1\rvert}

\newcommand{\bnorm}[1]{\bigl\lVert#1\bigr\rVert}

\hypersetup{
 colorlinks=true,
 citecolor=blue,
 linkcolor=blue,
 frenchlinks=false,
 pdfborder={0 0 0},
 naturalnames=false,
 hypertexnames=false,
 breaklinks}
\usepackage[capitalize,nameinlink]{cleveref}

\crefname{section}{Section}{Sections}
\crefname{subsection}{Subsection}{Subsections}
\crefname{observation}{Observation}{Observations}
\crefname{hypothesis}{Hypothesis}{Hypotheses}
\crefname{assumption}{Assumption}{Assumptions}
\crefname{claim}{Claim}{Claims}
\Crefname{figure}{Figure}{Figures}

\Crefname{figure}{Figure}{Figures}

\crefformat{equation}{\textup{#2(#1)#3}}
\crefrangeformat{equation}{\textup{#3(#1)#4--#5(#2)#6}}
\crefmultiformat{equation}{\textup{#2(#1)#3}}{ and \textup{#2(#1)#3}}
{, \textup{#2(#1)#3}}{, and \textup{#2(#1)#3}}
\crefrangemultiformat{equation}{\textup{#3(#1)#4--#5(#2)#6}}%
{ and \textup{#3(#1)#4--#5(#2)#6}}{, \textup{#3(#1)#4--#5(#2)#6}}%
{, and \textup{#3(#1)#4--#5(#2)#6}}

\Crefformat{equation}{#2Equation~\textup{(#1)}#3}
\Crefrangeformat{equation}{Equations~\textup{#3(#1)#4--#5(#2)#6}}
\Crefmultiformat{equation}{Equations~\textup{#2(#1)#3}}{ and \textup{#2(#1)#3}}
{, \textup{#2(#1)#3}}{, and \textup{#2(#1)#3}}
\Crefrangemultiformat{equation}{Equations~\textup{#3(#1)#4--#5(#2)#6}}%
{ and \textup{#3(#1)#4--#5(#2)#6}}{, \textup{#3(#1)#4--#5(#2)#6}}%
{, and \textup{#3(#1)#4--#5(#2)#6}}

\crefdefaultlabelformat{#2\textup{#1}#3}

\usepackage{fancyhdr}
\usepackage[margin=1in]{geometry}

\definecolor{dgreen}{rgb}{.0,.5,.0}

\definecolor{dblue}{rgb}{.0,.0,.6}
\definecolor{mblue}{rgb}{.0,.0,.8}
\definecolor{dred}{rgb}{.7,.0,.0}
\definecolor{mred}{rgb}{.8,.0,.0}
\definecolor{dmagenta}{rgb}{.4,.1,.5}
\definecolor{dgreen}{rgb}{.0,.4,.0}

\begin{document}

\TITLE{On Learning the \texorpdfstring{$c\mu$}{} Rule in Single\\ and Parallel Server Networks}
\RUNTITLE{On Learning the \texorpdfstring{$c\mu$}{} Rule in Single and Parallel Server Networks}
\RUNAUTHOR{Krishnasamy, Arapostathis, Johari, and Shakkottai}

\ARTICLEAUTHORS{
\AUTHOR{Subhashini Krishnasamy$^\dag$, Ari Arapostathis$^\ddag$,
Ramesh Johari$^\ast$, and Sanjay Shakkottai$^\P$}
\AFF{$^\dag\,$Tata Institute of Fundamental Research,
$^\ddag\,^\P\,$The University of Texas at Austin, $^\ast\,$Stanford University}
\EMAIL{subhashini.kb@utexas.edu, ari@ece.utexas.edu, rjohari@stanford.edu, \\
shakkott@austin.utexas.edu}
}

\ABSTRACT{\textbf{Abstract}.
We consider learning-based variants of the $c \mu$ rule
for scheduling in single and parallel server settings of multi-class queueing systems.

In the  single server setting, the $c \mu$ rule is known to minimize the
expected holding-cost (weighted queue-lengths summed over
classes and a fixed time horizon). We focus on the problem where the service rates
$\mu$ are unknown with the {\em holding-cost
regret} (regret against the $c \mu$ rule with known $\mu$) as our objective.
We show that the greedy algorithm that uses empirically learned service rates results in a {\em constant} holding-cost regret (the regret is independent of the time horizon). This {\em free exploration} can be explained in the single server setting by the fact that any work-conserving policy obtains the same number of samples in a busy cycle. 

In the parallel server setting, we show that the $c \mu$ rule may result in unstable queues, even for arrival rates within the capacity region. We then present sufficient conditions for geometric ergodicity under the $c \mu$ rule.
Using these results, we propose an almost greedy algorithm that explores only when the number of samples falls below a threshold.  We show that this algorithm delivers constant holding-cost regret because a free exploration condition is eventually satisfied.
%
%
}

\MSCCLASS{68M20, 93E35}


\maketitle

\section{Introduction.}

We consider a canonical scheduling problem in a discrete-time, multi-class, multi-server parallel server queueing system.  
In particular, we consider a system with $U$ distinct queues, and $K$ distinct servers.
Each queue corresponds to a different class of arrivals; arrivals queue $i$ are Bernoulli($\lambda_i$), i.i.d across time.
Service rates $\mu_{ij}$ are heterogeneous across every pair of queue $i$ and server $j$ (i.e., a ``link'').
At each time step, a central scheduler may match at most one queue to each server.
Services are also Bernoulli; thus jobs may fail to be served when matched, and in this case the policy is allowed to choose a different server for the same job in subsequent time step(s).
Jobs in queue $i$ incur a holding cost $c_i$ per time step spent waiting for service.
Letting $Q_i(t)$ denote the queue length of queue $i$ at time $t$, the performance measure of interest up to time $T$ is the cumulative expected holding cost incurred  up to time $T$: 
\[ \sum_{t=1}^T \sum_{i \in [U]} c_i \EE\left[ Q_i(t) \right]. \]
(All our analysis extends to the case where the objective of interest is a time-discounted cost, i.e., where the $t$'th term is scaled by $\beta^t$, where the discount factor satisfies $0 < \beta < 1$.)

Our emphasis in this paper is on solving this problem when the link service rates are {\em a priori unknown}; the scheduler only learns the link service rates by matching queues to servers, and observing the outcomes.
We use as our benchmark the {\em $c\mu$ rule} for scheduling, when link service rates are known.
The $c\mu$ rule operates as follows: at each time step, each link from a nonempty queue $i$ to server $j$ is given a weight $c_i \mu_{ij}$; all other links are given weight zero.
The server then chooses a maximum weight matching on the resulting graph as the schedule for that time step.
It is well known that when there is only a single server, this rule delivers the optimal expected holding cost among all feasible scheduling policies.
Further, there has been extensive analysis of the performance and optimality properties of this rule even in multiple server settings.  (See related work below.)

When service rates are unknown, we measure the performance of any policy using (expected) {\em regret} at $T$: this is the expected difference between the cumulative cost of the policy, and the cumulative cost of the $c\mu$ rule.
Our goal is to characterize policies that minimize regret.  
In typical learning problems such as the stochastic multiarmed bandit (MAB) problem, optimal policies must resolve an {\em exploration-exploitation} tradeoff.
In particular, in order to minimize regret, the policy must invest effort to learn about unknown actions, some of which may later prove to be suboptimal---and thus incur regret in the process.
In such settings, any optimal policy incurs regret that increases without bound as $T \to \infty$; for example, for the standard MAB problem, it is well known that optimal regret scales as $O(\ln T)$  \cite{lai1985asymptotically, auer2002finite, agrawal2011analysis}.

In this paper, we show a striking result: in a wide range of settings, the {\em empirical} $c\mu$ rule---i.e., the $c\mu$ rule applied using the current estimates of the mean service rates---is regret optimal, and further, the resulting optimal regret is bounded by a {\em constant} independent of $T$.
Thus, in such settings there is no tradeoff between exploration and exploitation.
The scheduler can simply execute the optimal schedule given its current best estimate of the services rates of the links.
In other words, the empirical $c\mu$ rule benefits from {\em free exploration}.

We make three main contributions: (1) regret analysis of the empirical $c\mu$ rule in the single server setting; (2) stability analysis of the $c\mu$ rule in the multi-server setting; and (3) subsequent regret analysis of the empirical $c\mu$ rule in the multi-server setting.  We summarize these contributions below.

\begin{enumerate}
\item {\bf Learning in the single-server setting}.  We begin our analysis by focusing on the single-server setting, where the $c\mu$ rule is known to be optimal on any finite time horizon.
This setting admits a particularly elegant analysis, due to the following two observations.
First, the empirical $c\mu$ rule is {\em work-conserving}, as is the benchmark $c\mu$ rule with known service rates.
Second, all work-conserving scheduling policies have the property that they induce the {\em same} busy period distribution on the queueing system.
Using this observation, we can couple the empirical $c\mu$ rule to the $c\mu$ rule with known service rates, and  divide our analysis into epochs defined by busy periods.
At the end of any busy period, all queue lengths are identical in both systems: namely, zero.
We show that after a sufficiently large number of busy periods have elapsed (namely, $O(\ln T)$),
with high probability the empirical $c\mu$ rule has sufficient knowledge of each arm that it exactly matches the $c\mu$ rule going forward.
Finally, we use the fact that any work-conserving policy induces a queue-length process that is geometrically ergodic to show that the expected regret is bounded by a constant.

\item {\bf An interlude: Stability in the multi-server setting.}  Next, we turn our attention to regret analysis in the setting of multiple servers.
Here, however, we face a challenge: in contrast to the single-server setting, where the $c\mu$ rule is known to be optimal, with multiple servers the $c\mu$ rule may not even be {\em stabilizing}, despite the availability  of sufficient service capacity.
Further, somewhat surprisingly there are no known general results in the literature on stability of the parallel server $c\mu$ rule.
In order to carry out regret analysis, of course, we require such conditions; therefore we develop them for our analysis.  These results are of independent interest.

We provide three results on stability.
First, we construct a class of examples that demonstrate that the rule need not be stabilizing.
Second, we develop a general condition for stability of the $c\mu$ rule on a particular class of queueing networks, where the rule takes the form of a {\em hierarchical static priority rule}.
Informally, these are networks where the configuration of service rates and costs is such that a priority structure among the queues can be embedded in a hierarchical graph.
In particular, we show for these systems that stability is equivalent to geometric ergodicity of the resulting queue-length process.
This condition is not directly over model primitives; thus in our third result we provide a stronger sufficient condition for geometric ergodicity of the $c\mu$ rule that can be directly checked on model primitives for a generic scheduling problem.
We show a number of network configurations for which this condition holds.

\item {\bf Learning in the multi-server setting.}  Having determined a sufficient condition for stability, we turn our attention to learning in the multi-server setting.
We show that for problem instances where the $c\mu$ rule with known service rates yields a geometrically ergodic queue length process, the empirical $c\mu$ rule yields a difference in queue lengths with the benchmark that decays at least polynomially with time.
As in the single server setting, this again results in $O(1)$ regret, following two insights that parallel our analysis of the single-server setting: 
first, that the system eventually reaches a state of ``free exploration''; and second, that the tails of the busy period can be shown to sufficiently light.
\end{enumerate}

\subsection{Related work.}  Many variants of the dynamic stochastic scheduling problem, for both discrete and continuous time queueing networks, have been long studied \cite{nino2008stochastic,mahajan2008multi}. Conventionally, it has been studied in the Markov decision process framework where it is assumed that the service rates are known a priori, and the proposed solution is usually an {\em index type} policy that schedules non-empty queues according to a static priority order based on the mean service time and holding-costs. The simplest variant of the problem is that of a multi-class single-server system for which the $c\mu$ rule has been shown to be optimal in different settings \cite{cox1961queues, buyukkoc1985cmu, hirayama1989further}. Klimov \cite{klimov1975time-sharing} extended the $c\mu$ rule to multi-class single-server systems with Bernoulli feedback. Van Mieghem \cite{van1995dynamic} studies the case of convex costs for a $G/G/1$ queue and proves the asymptotic optimality of the generalized $c\mu$ rule in heavy traffic. Ansell et al.\ \cite{ansell2003whittle} develop the Whittle's index rule for an $M/M/1$ system with convex holding-costs.

The works in \cite{harrison1998heavy, bell2001dynamic} study a simple parallel server model---the N-network, which is a two queue, two server model with one flexible and one dedicated server---and propose policies that achieve asymptotic optimality in heavy traffic. Glazebrook and Mora \cite{glazebrook2001parallel} consider the parallel server system with multiple homogeneous servers and propose an index rule that is optimal as arrival rates approach system capacity. Lott and Teneketzis \cite{lott2000optimality} also study the parallel server system with multiple homogeneous servers and derive sufficient conditions to guarantee the optimality of an index policy. Mandelbaum and Stolyar \cite{mandelbaum2004scheduling} study the continuous time parallel server system with non-homogeneous servers and convex costs. They prove the asymptotic optimality of the generalized $c\mu$ rule in heavy traffic. Among the above papers, only \cite{cox1961queues, buyukkoc1985cmu, hirayama1989further, lott2000optimality} consider the holding-cost across a finite horizon. The rest have their objective as the infinite horizon discounted
and/or average costs. Our work provides results for both the finite horizon discounted cost and finite horizon total cost problems.

Another framework in which the problem can be studied is the stochastic multi-armed bandit problem, where the aim is to minimize the regret in finite time. Traditional work in the space of MAB problems focuses on the exploration-exploitation tradeoff and investigates various exploration strategies to achieve optimal regret \cite{lai1985asymptotically, auer2002finite, agrawal2011analysis}. More recently, exploration-free or greedy algorithms have been studied and shown to be effective in a few contexts. \cite{mersereau2009structured} studies the linear bandit problem in the Bayesian setting and shows asymptotic optimality of a greedy algorithm with respect to the known prior. For a variant of the linear contextual bandits, Bastani et al. \cite{bastani2017exploiting} propose to reduce exploration by dynamically deciding to incorporate exploration only when it is clear that the greedy strategy is performing poorly. For a slightly different variant of the linear contextual bandits, Kannan et al. \cite{kannan2018smoothed} show that perturbing the context randomly and dynamically can give non-trivial regret bounds for the greedy algorithm with some initial training. In a similar vein, our work proposes to reduce exploration through a conditional-exploration strategy. We show that this policy eventually transforms into a purely greedy strategy because the system naturally provides free exploration.

\subsection{Organization of the paper.}
We describe the queueing model and main objective of this work in \cref{S2}.
In \cref{S4}, we present the analysis for the single server system. In \cref{S3}, we show that the stability region for the $c\mu$ rule is a strict subset of the capacity region and give sufficient conditions for geometric ergodicity under the $c\mu$ rule.
In \cref{sec:parallel-server}, we extend the analysis presented in \cref{S4} to parallel server systems and show constant order regret when the system is geometrically ergodic under the $c\mu$ rule.
Appendix~\ref{S6} is devoted to the study of a special class of scheduling rules called hierarchical rules, for which we exhibit a recursive procedure which verifies geometric ergodicity
from the system parameters.
The more technical proofs are organized in Appendices~\ref{AppA}--\ref{sec:dist-1x2}.

\section{Problem Setting.}\label{S2}

We describe the model, the objective, and the $c\mu$ rule.
\subsection{Parallel server system with linear costs.}
Consider a discrete-time parallel server system with $U$ queues (indexed by $i \in [U]$) and $K$  servers (indexed by $j \in [K]$). Jobs arrive to queue $i$ according to a Bernoulli process with rate $\lambda_i$ independent of other events. Denote the joint arrival process by $Bernoulli(\bm{\lambda})$. At any time, a server can be assigned only to a single job and vice-versa; however, multiple servers are allowed to be assigned to different jobs in the same queue.
For convenience of exposition, we assume that jobs are assigned according to FCFS. At any time, the probability that a job from queue $i$ assigned to server $j$ is successfully served is $\mu_{i,j}$ independent of all other events.
We denote this joint  service distribution by $Bernoulli(\bm{\mu})$. Jobs that are not successfully served remain in the queue and can be reassigned to any server in subsequent time-slots. The queues have infinite capacity, and $c_i$ denotes the waiting cost per job per time-slot for queue $i$.  For this system, a \emph{scheduling rule} is defined as one that decides, at the beginning of every time-slot, the assignment of servers to queues. It is assumed that
\begin{enumerate}
\item[(i)] the outcome of an assignment is not known in advance, i.e., in any time-slot, whether or not a scheduled job is served successfully can be observed only at the end of the time-slot;
\item[(ii)] the waiting cost per job is known for all the queues.
\end{enumerate}
We study the learning variant of the problem, and therefore make the additional assumption that
\begin{enumerate}
\item[(iii)] the arrival rates and success probabilities are unknown.
\end{enumerate}
For $T$ time-slots,  the expected total waiting cost in finite time  is given by
\begin{equation}\label{eq:total-cost}
J(T) \,\df\, \EE\left[ \sum_{t=1}^T \sum_{i=1}^U c_i Q_i(t) \right].
\end{equation}
Here $Q_i(t)$ is the queue-length of queue $i$ at the beginning of time-slot $t$, with the evolution dynamics given by the equation
\begin{equation*}
\bm{Q}(t+1) = \bigl( \bm{Q}(t) - \bm{S}(t) \bigr)^+
+ \bm{A}(t)	\quad \forall t \geq 0,
\end{equation*} 
where $\bm{A}(t)$ and $\bm{S}(t)$ are the arrival vector and allocated service vector respectively.
In \eqref{eq:total-cost}, the instantaneous waiting cost is a linear function of the queue-lengths.
\begin{definition}[Stability]
 For a Markov policy $\phi$, i.e., a scheduling rule that makes decisions in every time-slot based on the current queue-state, the system is said to be \emph{stable} under $\phi$ if the chain $\bm{Q}(t)$ is positive recurrent and 
\begin{equation*}
\int_{\ZZ_+^U}  \norm{q}_1\, \uppi ( d q)  < \infty,
\end{equation*} 
where $\uppi$ is its invariant distribution.
\end{definition} 

For a given service rate (success probability) matrix $\bm{\mu}$
and a Markov policy $\phi$, let the \emph{stability region}
$\mathscr{C}^{\phi}(\bm{\mu})$ be the set of all arrival rates for which the system is stable under $\phi$.
The \emph{capacity region} of the parallel server system with service rate
matrix $\bm{\mu}$ is given by
$\mathscr{C}(\bm{\mu}) \,\df\, \cup_{\phi} \mathscr{C}^{\phi}(\bm{\mu})$. The capacity region can be characterized by the class of static-split scheduling policies.
\begin{equation*}
\mathscr{C}(\bm{\mu}) = \bigl\lbrace \bm{\lambda}\,\colon \bm{\lambda} < \diag( \bm{\mu} \bm{M} ),\;  \bm{M}
\in \mathcal{P}_{K \times U}  \bigr\rbrace,
\end{equation*}
where $\mathcal{P}_{K \times U}$ is the set of all $K \times U$ right
stochastic matrices.

\subsection{The \texorpdfstring{$c\mu$}{} rule.}
In this paper, we focus on the $c\mu$ rule with linear costs for the parallel server system. This rule (see \cref{alg:c-mu}), which is a straightforward generalization of the single server $c\mu$ rule, allocates servers to jobs based on a priority rule determined by the product of the waiting cost and success probability.
\begin{algorithm}\label{Alg1}
  \caption{The $c\mu$ Algorithm with costs $\bm{c}$ and rates $\bm{\mu}$}
  \begin{algorithmic}
    \State At time $t$:
    \State Solve the following {\em max weight} optimization problem:
\begin{align*}
\text{maximize}\ \ \ & \sum_{i,j} c_i \mu_{i,j} x_{ij} \\
\text{subject to}\ \ \ & \sum_i x_{ij} \leq 1;\\
& \sum_j x_{ij} \leq Q_i(t);\\
& x_{ij} \in \{0, 1\}.\\
\end{align*}
\State Assign a job from queue $i$ to server $j$ if and only if $x_{ij} = 1$ in the resulting solution.
  \end{algorithmic}
    \label{alg:c-mu}
\end{algorithm}

For a single server system, and when the success probabilities for all the links are known a priori, it has been established that the $c\mu$ rule optimizes the expected total waiting cost over a finite time horizon \cite{buyukkoc1985cmu}.
For a parallel server system, there are no known algorithms that achieve optimal cost as in a single server system. 
For waiting costs that are strictly convex in queue-lengths, Mandelbaum and Stolyar \cite{mandelbaum2004scheduling} prove that, in heavy-traffic, the generalized $c\mu$ rule optimizes the instantaneous waiting cost asymptotically. 

In order for the $c\mu$ rule to be unambiguously defined, we impose
the assumption that
\begin{equation}\label{E-Delta}
\varDelta \df \min_{j,j' \in [K]}\, \min_{i, i' \in [U]}\,
\bigl\lbrace\card{c_i \mu_{i,j} - c_{i'} \mu_{i',j}}
\wedge\card{c_i \mu_{i,j} - c_{i} \mu_{i,j'}}\,\colon \mu_{i,j}\ne0\,,\,
i\ne i',\, j\ne j'\bigr\rbrace
\,>\,0\,.
\end{equation}

Our interest lies in designing scheduling algorithms that can mimic the  $c\mu$ rule in the absence of channel statistics. We evaluate an algorithm based on a finite time performance measure called \emph{regret}. Conventionally, in bandit literature, regret measures the difference in the performance objective  between 
an adaptive algorithm and a genie algorithm that has an a priori
knowledge of the system parameters. For our problem, the genie algorithm
applies the $c\mu$ rule at every step, using the service matrix $\bm{\mu}$.
Therefore, regret here is defined as the difference between total waiting costs
(given by equation~\eqref{eq:total-cost}) under the proposed algorithm and the
$c\mu$ algorithm.
For any given parameter set
$(\bm{\lambda}, \bm{\mu})$ such that
$\bm{\lambda} \in \mathscr{C}^{c\mu}(\bm{\mu})$, we study the asymptotic behavior
of regret as the time-period $T$ tends to infinity. 

\section{Learning the \texorpdfstring{$c\mu$}{} Rule---Single Server System.}
\label{S4}

We first consider the single server system in order to highlight a few key aspects of the problem. We later extend our discussion and results to the parallel-server case in \cref{sec:parallel-server}. For the single server system, we propose a natural `learning' extension of the $c \mu$ algorithm, which we refer to as the $c\hat{\mu}$ algorithm,
or $c\hat{\mu}$ rule.
This scheduling algorithm applies the $c \mu$ rule using empirical means
for $\bm\mu$ obtained from past observations as a surrogate for the actual success probabilities. Let us denote the queue-lengths  under the $c \hat{\mu}$ and $c\mu$ rules by $\bm{Q}$ and $\bm{Q^*}$ respectively. Further, we denote the regret of the $c\hat{\mu}$ algorithm by
\begin{equation*}
\Psi(T)\,\df\, J(T) - J^*(T),
\end{equation*}
where $J(T)$ and $J^*(T)$ are the respective expected total waiting costs.


We show that the queue-length error for the $c\hat{\mu}$ algorithm decays geometrically with time.
It then follows that the regret scales as a constant with increasing $T$ for any $\bm{\lambda} \in \mathscr{C}^{c\mu}(\bm{\mu})$.
It is interesting to observe that this scaling is achieved only by using the empirical means in every time-slot, without an explicit explore strategy.  Our results show that this scheduling policy delivers {\em free exploration} due to some unique properties of the single server system, as we describe further below (see \cref{prop:c-mu-hat}).
For single server systems, the definition in \cref{E-Delta} takes the form
$\varDelta \df \min_{i\ne j} \card{c_i \mu_i - c_j \mu_j} > 0$.

\begin{proposition}\label{prop:c-mu-hat}
For any $(\bm{\lambda}, \bm{\mu}, \bm{c}, \bm{Q}(0))$ such that $\bm{\lambda} \in \mathscr{C}^{c\mu}(\bm{\mu})$, there exist constants $C_0 > 0$ and $\rho \in (0,1)$ such that
$$\EE\left[ \bnorm{\mathbf{Q}(t) - \mathbf{Q}^*(t)}_1 \right] \le C_0 \rho^t$$ for any $t \in \NN$.
In particular, there exists a constant $C$ independent of $T$ such that the regret $\Psi(T)$ satisfies $\limsup_{T \to \infty} \Psi(T) = C$.
\end{proposition}

Before proving the result, we briefly outline the intuition.  The result relies on the following key observation.  

\begin{observation}
\label{obs:same-busy-cycle}
The distribution of busy cycles is the same for all work conserving scheduling
policies in a single server system.
\end{observation}

This can be confirmed by considering a stochastically equivalent system where,
for any $i \in [U]$, jobs arrive to queue $i$ with i.i.d.\ inter-arrival times
distributed as $Geom(\lambda_i)$ and i.i.d.\ service times distributed as $Geom(\mu_i)$.
In such a system, a scheduling algorithm only decides which part of the work
is completed in each time slot and therefore,
all work conserving algorithms give the same busy cycle.

We now see how \cref{obs:same-busy-cycle} can be used to prove
\cref{prop:c-mu-hat}. This observation implies that the $c\hat{\mu}$ and $c\mu$ systems have the same queue length (equal to $\bm{0}$) at the end of their common busy cycles.  In order for the estimated priority order by the $c\hat{\mu}$ algorithm to agree with $c\mu$, it needs sufficient number of samples for all the links. Since the number of samples for each queue at the end of a busy cycle is equal to the total work (in terms of service time) arrived to the queue, it is sufficient to consider the end of a busy cycle by which the system has seen at least $O (\log t)$ arrivals to every queue. Thus, every work conserving policy has the same number of samples for each of the links at the end of a busy cycle.  Finally, we exploit the fact that busy periods have geometrically decaying tails to show that as a consequence, the $c\hat{\mu}$ algorithm makes the same scheduling decision as the $c\mu$ rule after a random time $\tau$ that has finite expectation.   This argument is a clear example of {\em free exploration}, since there is no need to incorporate an explicit exploration strategy into the scheduling algorithm as long as it is work conserving.  

\proof{Proof of \cref{prop:c-mu-hat}.}
The crux of the proof lies in characterizing the random time $\tau$ after which the $c\hat{\mu}$ algorithm makes the same scheduling decision as the $c\mu$ rule in all future time-slots. In any time-slot $t$, the $c\hat{\mu}$ algorithm makes the same scheduling decision as the $c\mu$ rule if 
(i) $\bm{Q}(t)=\bm{Q}^*(t)$, and (ii) the estimated priority order agrees with the $c\mu$ rule at $t$. 





Our argument crucially relies on \cref{obs:same-busy-cycle}. 
We start by noting that the queue-length process under any work-conserving algorithm is geometrically ergodic and the busy cycle lengths have geometrically decaying tails. Specifically, there exist constants $r_1, r_2 > 1$ and $C_1 > 0$ such that the first hitting time
of the state $\bm q=\bm0$, denoted by $\uptau_{\bm0}$, satisfies
\begin{equation}
\label{eq:geom-ergod}
\EE\left[ r_2^{\uptau_{\bm0}} \right] \le C_1 r_1^{\norm{\mathbf{Q}(0)}_1}.
\end{equation}

To formalize the intuition in the paragraph preceding the proof, let $t' = \lfloor \frac{t}{k_0} \rfloor$ where $k_0 := \max\left\lbrace 1, \frac{U \log r_1 + \log r_2}{\log(\rho_1 r_2)} \right\rbrace$ for some $\rho_1 \in (0,1)$, and let $\tau(t')$ be  the end of the busy period that contains $t'$. Then from \cref{eq:geom-ergod}, using Markov's inequality, we have
\begin{equation}
\label{eq:geom-tail}
\PP\left[ \tau(t') \ge t - t' \right] \le \frac{C_1 r_1^{\norm{\mathbf{Q}(0)}_1 + U t'}}{r_2^{t - t'}} \le C_1 r_1^{\norm{\mathbf{Q}(0)}_1} \rho_1^t,
\end{equation}
where the second inequality follows by the definition of $k_0$.
Now, let $\hat{\mu}_{i}^{(n)}$ be the average number of successes in the first $n$ assignments of the server to queue $i$.
 Consider the following two events:
\begin{enumerate}
\item \event{ev:enough-samples}$\mathscr{E}_{\ref{ev:enough-samples}} \df \left\lbrace \sum_{l=1}^{t'} A_i(l) >  \nicefrac{\lambda_i t'}{2} \quad \forall i \in [U] \right\rbrace $,
\item \event{ev:accurate-est}$\mathscr{E}_{\ref{ev:accurate-est}} \df \left\lbrace c_i \card{\hat{\mu}_{i}^{(n)} - \mu_i} < \nicefrac{\varDelta}{2} \quad \forall n \geq \nicefrac{\lambda_i t'}{2}, \forall i \in [U] \right\rbrace$.
\end{enumerate}
Then, conditioned on
$\mathscr{E}_{\ref{ev:enough-samples}} \cap \mathscr{E}_{\ref{ev:accurate-est}}$,
the $c\hat{\mu}$ algorithm agrees with the $c\mu$ rule after $t'$, and
therefore its queue-length equals that of $c\mu$ after $t' + \tau(t')$.
Thus, given $\mathscr{E} = \mathscr{E}_{\ref{ev:enough-samples}} \cap \mathscr{E}_{\ref{ev:accurate-est}} \cap \{\tau(t') < t - t'\}$, we have $\bm{Q}(t) = \bm{Q}^*(t)$. 
It is easy to show the following using the Chernoff--Hoeffding bound for Bernoulli random variables.
\begin{equation}
\label{eq:rules-align}
\PP\left[ \mathscr{E}_{\ref{ev:enough-samples}}^c \right] + \PP\left[ \mathscr{E}_{\ref{ev:accurate-est}}^c \right] \leq C_2 \rho_2^t,
\end{equation}
 for some $C_2 > 0$ and $\rho_2 \in (0,1).$

Using bounds~\eqref{eq:geom-tail}, \eqref{eq:rules-align}, for any norm function $\norm{\cdot}$, we have
\begin{align*}
\EE\left[ \bnorm{\mathbf{Q}(t) - \mathbf{Q}^*(t)} \right] & = \EE\left[ \bnorm{\mathbf{Q}(t) - \mathbf{Q}^*(t)} \mathds{1}_{\mathscr{E}^c} \right]	\\[3pt]
& \le \left( \norm{\mathbf{Q}(0) + t \mathbf{1}} \right) \PP\left[ \mathscr{E}^c \right]	\\[3pt]
& \le \left( \norm{\mathbf{Q}(0) + t \mathbf{1}} \right) \left( C_1 r_1^{\norm{\mathbf{Q}(0)}_1} \rho_1^t + C_2 \rho_2^t \right)	\\
& \leq C_0 \rho^t,
\end{align*}
for some $C_0 > 0$ and $\rho \in (0,1)$.
This also shows that the regret $\Psi(T)$ scales as $O(1)$ with $T$.
\begin{align*}
\Psi(T) = \sum_{t=1}^T \EE\left[ \sum_{i=1}^U c_i \bigl( Q_i(t) - Q^*_i(t)\bigr) \right] &\le \left( \max_{i\in [U]} c_i \right) \sum_{t=1}^T \EE\left[ \bnorm{\mathbf{Q}(t) - \mathbf{Q}^*(t)}_1 \right] \\[5pt]
&\leq \left( \max_{i\in [U]} c_i \right) \frac{C_0 \rho}{1 - \rho}\,.\qquad\Halmos
\end{align*}
\endproof

\begin{remark}
\label{rem:ext1}
Note that an $O(1)$ scaling with $T$ also holds for regret with discounted cost for any discount factor.
\end{remark}

\section{Stability of the \texorpdfstring{$c\mu$}{} Rule for Parallel Server Systems.}
\label{S3}
As for the single server system, we are interested in upper bounds on regret for the parallel server system.
 In  the proof of \cref{prop:c-mu-hat}, we crucially used the property of identically distributed busy cycles over work conserving policies. Note that, in this case, the stability region of $c\mu$ rule (or any work conserving policy) is the entire capacity region, and the busy cycles have exponentially decaying tails for any arrival rate in this region. 

In this section, we show for the parallel server system that the $c\mu$ rule (which is based on linear costs) does not necessarily ensure stability for all arrival rates in the capacity region.
In particular, it is not throughput optimal for a general parallel server system.  In \cref{sec:cmu-geom-ergod}, we characterize a subset of the stability region of $c\mu$ rule for which the busy cycles have  exponentially decaying tails.

\subsection{Instability of the $c\mu$ rule in the general case.}
\label{ssec:instability}

As defined in \cref{Alg1}, the $c\mu$ rule allocates server $j$ to a job in the queue that maximizes $c_i \mu_{i, j}$. We show that such a \emph{static priority} policy, which prioritizes queues irrespective of their queue-lengths (other than their being non-empty) could be detrimental to the stability of the system.
For e.g., in any $2 \times 2$ system with
$c_2 \mu_{2,1} < c_1 \mu_{1,1},\; c_2 \mu_{2,2} < c_1 \mu_{1,2}$,
the $c\mu$ rule prioritizes $Q_1$ over $Q_2$ for allocation of both the servers, which results in service allocation
to $Q_2$ only when there are less than $2$ jobs in $Q_1$.
It is intuitively clear that such a policy is not stabilizing if the arrival
rate of $Q_2$ is larger than the service rate that this policy can allocate to $Q_2$.
We formalize this in the theorem below, where we characterize the set of arrival rates outside the stability region of the $c\mu$ rule for a class of $2 \times 2$ systems.

\begin{theorem}\label{thm:cmu-instability}
For any $2 \times 2$ system with service rates $\bm{\mu}$, costs
$\mathbf{c}$,  and arrival rates $\bm\lambda$ satisfying
\begin{equation}
\label{eq:2x2-structure}
c_2 \mu_{2,1} < c_1 \mu_{1,1},\quad c_2 \mu_{2,2} < c_1 \mu_{1,2},
\quad \mu_{1,2} < \mu_{1,1},
\end{equation}
and
\begin{equation}
\label{eq:2x2-unstable}
\lambda_1 < \mu_{1,1} + \mu_{1,2},\quad \lambda_2 > \uppi_1(0) \mu_{2,1} + \uppi_1(\{0,1\}) \mu_{2,2},
\end{equation}
where $\uppi_1$ is the stationary distribution of the Markov chain $\{Q_1(l)\}_{l>0}$, there exist positive constants $b_1, b_2, t_0$ depending on
$(\bm{\lambda}, \bm{\mu})$ such that
$$\PP\left[ Q_2(t) < b_2 t \right] \leq \exp\left( -b_1 t \right)
\qquad\forall\,t \geq t_0.$$
\end{theorem}
 
It is easy to construct an example of a $2 \times 2$ system with parameters
$\bm{\lambda}, \bm{\mu}, \bm{c}$ satisfying \cref{eq:2x2-structure,eq:2x2-unstable}
and with $\bm{\lambda} \in \mathscr{C}(\bm{\mu})$.
This shows that for $2 \times 2$ systems,
the stability region of the $c\mu$ rule is,
in general, a strict subset of the capacity region.
Below, we give such an example:
\begin{example}
Pick any $\mu_{1,1}, \mu_{1,2}, \lambda_1 \in (0,1)$ such that $\mu_{1,1} > \mu_{1,2}$, and $\mu_{1,1} > \lambda_1$. For this choice of  $\mu_{1,1}, \mu_{1,2}, \lambda_1$, let  $\uppi_1$ be the stationary distribution of $Q_1$ when served by both servers. Now pick $\mu_{2,1}, \mu_{2,2}, \lambda_2 \in (0,1)$ such that $\mu_{2,2} > \lambda_2 > \uppi_1(0) \mu_{2,1} + \uppi_1(\{0,1\}) \mu_{2,2}$.
Next, choose $c_1, c_2 > 0$ such that $c_2 \mu_{2,1} < c_1 \mu_{1,1}$, and $c_2 \mu_{2,2} < c_1 \mu_{1,2}$. Clearly, $\bm{\lambda} \in \mathscr{C}(\bm{\mu})$,
 since $\mu_{1,1} > \lambda_1$ and $\mu_{2,2} > \lambda_2$.
Thus, since the system parameters satisfy \cref{eq:2x2-structure,eq:2x2-unstable},
it follows by \cref{thm:cmu-instability} that
$\bm{\lambda} \notin \mathscr{C}^{c\mu}(\bm{\mu})$.
\end{example}

The criterion for instability in \cref{thm:cmu-instability} is rather sharp,
and this is evidenced by the following result.

\begin{theorem}\label{T3.2}
Any $2\times 2$ system with service rates $\bm{\mu}$ and costs
$\bm{c}$ satisfying \eqref{eq:2x2-structure} is stable under the
$c\mu$ rule if and only if
\begin{equation}\label{ET3.2A}
\lambda_1 < \mu_{1,1} + \mu_{1,2},\quad\text{and\ \ } \lambda_2
< \uppi_1(0) \mu_{2,1} + \uppi_1(\{0,1\}) \mu_{2,2},
\end{equation}
where $\uppi_1$ is the stationary distribution of the Markov chain $\{Q_1(l)\}_{l>0}$.
In addition, \eqref{ET3.2A} implies that the queueing process
$\bm{Q}(t)=\bigl(Q_1(t),Q_2(t)\bigr)$
is geometrically ergodic under the $c\mu$ rule.
In particular, there exists a function $\Lyap \colon\ZZ^2_+\to[1,\infty)$,  such that
$\E^{\epsilon_1 \norm{\bm q}_1}\le\Lyap(q)\le \E^{\epsilon_2 \norm{\bm q}_1}$
for some positive constants $\epsilon_1$ and $\epsilon_2$,
and constants $\rho\in(0,1)$, $B > 0$ and a finite set $\sB$, such that
$P\Lyap(\bm q)\le B \Ind_{\sB} + \rho \Lyap(\bm q)$,
where $P$ denotes the transition kernel of the chain $\bm{Q}(t)$.
It is well known that this implies that there exist constants $\gamma>0$ and $C>0$ such that  $\uptau$, the first hitting time of the state $\bm q=\bm0$, satisfies
\begin{equation}\label{eq:exp-moment-busy-cycle}
\EE\bigl[(\nicefrac{\rho}{2})^{\uptau} \bigr]\le C \E^{\gamma\norm{\bm{Q}(0)}_1}\qquad
\forall\, \bm{Q}(0)\in\ZZ_+^2\,.
\end{equation}
\end{theorem}

%
%
%

The proofs of \cref{thm:cmu-instability,T3.2} can be found in Appendix~\ref{AppA}.
%

\subsection{Sufficient conditions for geometric ergodicity of the \texorpdfstring{$c\mu$}{} system.} 
\label{sec:cmu-geom-ergod}
We now obtain sufficient conditions for the busy cycles to have exponentially decaying tails in terms of
the parameters $(\bm\lambda,\bm\mu)$. This condition, in particular, implies that the queue-length process is geometrically ergodic.

Let $\mathscr{Q}_l := \{\bm{q} \in \ZZ_+^U : \norm{\bm{q}}_1 = l\}$ for $l \in \ZZ_+$. For any $\bm{q} \in \mathscr{Q}_K$, let $R_i(\bm{q})$ denote the total service rate assigned by the $c\mu$ rule to queue $i$ when the queue-state is $\bm{q}$.
If
\begin{equation}\label{eq:feasibility}
 \bm{\lambda} \cdot \bm{\alpha} < \min_{\bm{q} \in \mathscr{Q}_K} (\bm{R}(\bm{q}) \cdot \bm{\alpha}),
\end{equation}
for some $\bm{\alpha} > 0\,,\ \bm{\alpha} \in \mathcal{P}_U$, where $\mathcal{P}_U$ is the probability simplex in $\RR^U$, then we can construct an appropriate Lyapunov function for which the one-step drift given by the $c\mu$ algorithm is negative outside a finite set.
This enables us to show the following tail probability bound for the busy period of the $c\mu$ system.

\begin{lemma}\label{lem:conc-busy-period}
Let $\uptau_{\bm{0}}$ denote the first hitting time of the
state $\bm q=\bm0$ under the $c\mu$ rule.
 If Condition \eqref{eq:feasibility} is true for some $\bm{\alpha} > 0\,,\ \bm{\alpha} \in \mathcal{P}_U$, then
there exist constants \const{const:3}$C_{\ref{const:3}}$, \const{const:4}$C_{\ref{const:4}}$, \const{const:5}$C_{\ref{const:5}}$ such that, for any $\kappa \in \RR$, 
\begin{equation}
\label{eq:conc-busy-period}
\PP\Bigl[ \uptau_{\bm{0}} > C_{\ref{const:3}} \kappa \log t + C_{\ref{const:4}} \bm{Q}(0) \cdot \bm{\alpha} + C_{\ref{const:5}} \Bigr] \le \frac{1}{t^{\kappa}}.
\end{equation}
\end{lemma}
Details of the proof of this lemma are given in
Appendix~\ref{AppB}.

Below, we explicitly derive sufficient conditions given by \eqref{eq:feasibility} for a couple of examples. Further, for the case of  the N-network in \cref{ex:N-nw} (which is a special case of the $2 \times 2$ network in \cref{T3.2}), we compare it with the stability region. 

%
\begin{example}
  Consider the $2 \times K$ example where queue $1$ has priority over
  queue $2$ for all $K$ servers. Without loss of
    generality, let $\mu_{1,1} > \mu_{1,2} > \dotsb > \mu_{1,K}$, and
    let $\uppi$ be the stationary distribution of $Q_1$ (note that,
    in every time-slot, service offered to $Q_1$ is independent of the
    current queue-length of $Q_2$). Then, the stability region
    $\mathscr{C}^{c\mu}$ is given by
\begin{equation*}
\lambda_1 < \sum_{k \in [K]} \mu_{1,k},\quad\text{and\ \ }
\lambda_2 < \sum_{q=0}^{K-1} \uppi(q) \sum_{k=q+1}^K \mu_{2,k}.
\end{equation*}
We now obtain a subset of the region \eqref{eq:feasibility} by
    choosing specific values of $\bm{\alpha}$.
For $\bm{\alpha}=(1, 1)$, \eqref{eq:feasibility} is satisfied if
\begin{equation*}
 \lambda_1 + \lambda_2 < \sum_{k \in [K]} \mu_{2,k}.
\end{equation*}
To see this, note that for any $1 \leq q \leq K$, we have
$$R_1(q, K-q) + R_2(q, K-q) \geq R_1(q-1, K-q+1) + R_2(q-1, K-q+1).$$
Therefore,
\begin{equation*}
\min_{\bm{q} \in \mathscr{Q}_K} (\bm{R}(\bm{q}) \cdot \bm{\alpha}) = R_2(0, K) = \sum_{k \in [K]} \mu_{2,k},
\end{equation*}
which shows that the region given by \eqref{eq:feasibility} contains
$\lambda_1 + \lambda_2 < \sum_{k \in [K]} \mu_{2,k}$.
\end{example}

\begin{example}\label{ex:N-nw}
Consider the N-network, i.e., a $2 \times 2$ system with $\mu_{2,1} =
0$, and let the first queue have higher priority according to the
$c\mu$ rule, i.e., $c_2 \mu_{2,2} < c_1 \mu_{1,2}$. This is a special case of the $2 \times 2$ system in \cref{T3.2}. Let $\uppi$ be
the stationary distribution of $Q_1$. A closed form expression for
$\uppi$ can be found in Appendix~\ref{sec:dist-1x2}. Thus, for this
system, we can determine the stability region analytically through \eqref{ET3.2A}. Moreover, as seen in \cref{T3.2}, we have geometric ergodicity
in all of the stability region $\mathscr{C}^{c\mu}(\bm{\mu})$.
Below, we compare the region given by
\eqref{eq:feasibility} with $\mathscr{C}^{c\mu}(\bm{\mu})$.

\paragraph{Case 1: $\mu_{1,1} \geq \mu_{1,2}$} -- Server $1$ is allocated to Queue $1$ when it has only a single job in its queue.
 In this case, as discussed above, the stability region $\mathscr{C}^{c\mu}(\bm{\mu})$ is given by
\begin{equation*}
\lambda_1 < \mu_{1,1} + \mu_{1,2},\quad\text{and\ \ }
\lambda_2 < \left( \uppi(\{0,1\}) \right) \mu_{2,2}\,,
\end{equation*}
whereas, Condition~\eqref{eq:feasibility} is equivalent to
$$\lambda_2 < \left( 1 - \frac{\lambda_1}{\mu_{1,1} + \mu_{1,2}}
\right) \mu_{2,2}.$$ This is the stability region of a $2 \times 1$
$c\mu$ system where the server has rates $\mu_{1,1} + \mu_{1,2}$ to
the first queue and $ \mu_{2,2}$ to the second queue.

\paragraph{Case 2: $\mu_{1,1} \leq \mu_{1,2}$} -- Server $2$ is allocated to Queue $1$ when it has only a single job in its queue.
In this case, the stability region $\mathscr{C}^{c\mu}(\bm{\mu})$ is given by
\begin{equation*}
\lambda_1 < \mu_{1,1} + \mu_{1,2},\quad\text{and\ \ }
\lambda_2 < \uppi(0)\mu_{2,2}\,,
\end{equation*}
whereas, Condition~\eqref{eq:feasibility} is equivalent
to $$\lambda_2 < \left( 1 - \frac{\lambda_1}{\mu_{1,2}} \right)^+
\mu_{2,2}\,.$$ 
%
In this example,  while Condition~\eqref{eq:feasibility} does not cover the entire stability region, the region it covers is ``close'' to the stability region in some limiting regimes. 
For example, in Case 1, when $\mu_{1,1} \gg \max\{\mu_{1,2}, \lambda_1\}$, we can show
that $$\uppi(\{0,1\}) \approx  \uppi(0) \approx 1 -
\frac{\lambda_1}{\mu_{1,1}} \approx 1 - \frac{\lambda_1}{\mu_{1,1} +
  \mu_{1,2}}.$$  
Similarly, in Case 2, when $\mu_{1,2} \gg \mu_{1,1}$, $\mu_{1,2} > \lambda_1$, and $
\frac{\mu_{1,2}}{1- \mu_{1,2}} \gg \frac{4\lambda_1
  \mu_{1,1}}{1-\lambda_1}$, we can show that $$\uppi(0) \approx 1 -
\frac{\lambda_1}{\mu_{1,2}}.$$  
\end{example}

\section{Learning the \texorpdfstring{$c \mu$}{} Rule---Parallel Server System.}
\label{sec:parallel-server}
\subsection{The \texorpdfstring{$c\hat{\mu}$}{} algorithm.}
We now propose a learning extension of the $c\mu$ rule for the parallel server system. 
Recall that the number of samples for a link in any time-slot is the number of times it has been scheduled before that time-slot. For the single server system, a sufficient number of samples can be ensured without explicit exploration due to the stabilizing property of work-conserving policies, all of which have the same busy periods. 
However, this property does not hold in general for the parallel server system, and thus, a straightforward extension of the $c\mu$ rule based on empirical means without explicit exploration may not obtain enough samples to learn the system. The following example shows how a naive extension of the $c\mu$ rule could fail to stabilize a $2 \times 2$ network.
\begin{example}
Consider a $2 \times 2$ network with service rates $\bm{\mu}$, costs
$\mathbf{c}$,  and arrival rates $\bm\lambda$ satisfying
\begin{equation*}
 \mu_{1,2} < \lambda_1 <  \mu_{1,1},\quad  \mu_{2,1} < \lambda_2 < \mu_{2,2},
\quad c_1 = c_2.
\end{equation*}
Clearly, this network is stable under the $c\mu$ rule. We show that, under the policy that does not explore and schedules according to the empirical estimates of the service rates, the queues have linear growth with positive probability. For any $i, j \in \{1,2\}, l \in \NN$, let $\Hat\mu_{i,j}^l$ be the empirical estimate of $\mu_{i,j}$ with $l$ samples. Let $\mathscr{E}$ be the event that $\Hat\mu_{1,1}^1 = \Hat\mu_{2,2}^1 = 0$ and $\Hat\mu_{1,2}^1 = \Hat\mu_{2,1}^1 = 1$.
Conditioned on the event $\mathscr{E}$ (which has a positive probability),
the $c\hat{\mu}$ algorithm
schedules only links $(1,2)$ and $(2,1)$ after obtaining the initial samples. Using Hoeffding's inequality, we can derive concentrations for the total number of arrivals to each of the queues and the total service offered by links $(1,2)$ and $(2,1)$ to show that there exist constants $t_0 \in \NN$, $b > 0$ such that $\PP\left[ \min \{Q_1(t), Q_2(t)\} > b t \quad \forall t > t_0 \right] > 0$.
\end{example}

As a solution to the above problem, we propose an algorithm that dynamically decides to explore if the number of samples falls below a threshold.
We refer to this as the $c\hat{\mu}$ algorithm for parallel server networks, and
define it in \cref{alg:c-mu-hat} below.

\subsubsection{Dynamic explore---conditional \texorpdfstring{$\epsilon$}{}-greedy.}
In each time-slot, the algorithm explores conditionally based on the number of samples, i.e., uses an $\epsilon$-greedy policy if the minimum number of samples over all links is below some threshold.
More specifically, let:
\begin{enumerate}
\item $\mathscr{E}$ be a collection of $U$ assignments such that their union covers the complete bipartite graph;
\item $N_{i,j}(t)$ be the number of samples of 
link $(i,j)$ at time $t$;
\item $N_{min}(t) = \min_{i, j} N_{i,j}(t)$;
\item $\Upsilon (t) = \max \left\lbrace 1, 2 \log^3(t-1) \right\rbrace$;
\item $\hat{\bm{\mu}}(t)$ be the estimated rate matrix at time $t$.
\end{enumerate}
At time $t$, 
if $N_{\mathsf{min}}(t) < \Upsilon (t)$, the algorithm decides to explore with probability $\nicefrac{\mathrm{poly}(\log t)}{t}$, otherwise it follows the $c\mu$ rule using the estimated rate matrix  $\hat{\bm{\mu}}(t)$.
\begin{algorithm}
  \caption{The $c\hat{\mu}$ algorithm for parallel server networks}
  \begin{algorithmic}
    \State At time $t$, 
    \State $\varepsilon(t) \leftarrow \ind{N_{\mathsf{min}}(t) < \Upsilon (t)},$ 
    \State $\mathsf{B}(t) \leftarrow $ independent Bernoulli sample of mean $\min\{1,3U\frac{\log^2t}{t}\}.$
    \If {$\varepsilon (t) \land \mathsf{B}(t) = 1$}
    \State \textit{Explore:} 
    Schedule from $\mathscr{E}$ uniformly at random.
    \Else
    \State \textit{Exploit:}
    Schedule according to the $c\mu$ rule with parameters $\hat{\bm{\mu}}(t)$.
    \EndIf 
  \end{algorithmic}
    \label{alg:c-mu-hat}
\end{algorithm}

\subsection{\texorpdfstring{$O(1)$}{} regret for the \texorpdfstring{$c\hat{\mu}$}{} algorithm.}
In \cref{thm:c-mu-hat}, we prove a regret bound that scales as a constant with increasing $T$ for a subset of the capacity region. This subset is given by the region in which the $c\mu$ algorithm achieves exponentially decaying busy cycles. 
In the theorem which follows,
we show that the queue-length error for the $c\hat{\mu}$ algorithm decays super-polynomially with time if \eqref{eq:conc-busy-period} is satisfied. Again, as in the single server system, this translates to an $O(1)$ regret. 
\begin{theorem}\label{thm:c-mu-hat}
For any $(\bm{\lambda}, \bm{\mu}, \bm{c}, \bm{Q}(0))$ such that \eqref{eq:conc-busy-period} is satisfied, we have
$$\lim_{t \to \infty} t^k  \EE\left[ \bnorm{\mathbf{Q}(t) - \mathbf{Q}^*(t)}_1 \right] = 0$$ for any $k \in \NN$.
In particular, there exists a constant $C$ independent of $T$ such that the regret $\Psi(T)$ satisfies $\limsup_{T \to \infty} \Psi(T) = C$.
\end{theorem}
As for the single server system, the main idea in proving \cref{thm:c-mu-hat} is to characterize the coupling time of the queue-lengths of the actual and the genie systems. More specifically, we show that the queue-length of the $c\hat{\mu}$ system at time $t$
does not exceed that of the genie system with probability
$O(\nicefrac{1}{\mathrm{poly}(t)})$. For this, we first show in \cref{lem:correct-order} that the $c\hat{\mu}$ algorithm obtains sufficient number of samples due to its conditional explore policy, thus enabling the algorithm
to agree with the $c\mu$ rule in its exploit phase
after time $\sqrt{t}$. In turn, this ensures exponentially decaying tails for the busy cycles after time $\sqrt{t}$ according to \cref{lem:conc-busy-period}.

This concentration for the busy cycles can be used to further show that
the following two `events' occur with polynomially high probability: 
\begin{enumerate}
\item That the algorithm does not need to explore in the latter half of $(0,t]$
(\cref{lem:no-explore}). 
This can be explained as follows: whenever the system hits the zero state,
there is a positive probability that only selective queues are non-empty in the subsequent time-slots. Therefore, for any work-conserving algorithm, every link has a positive probability of being scheduled at the beginning of a new busy cycle. If the algorithm stabilizes the system well enough to ensure that it hits the zero state regularly, then it obtains a sufficient number of samples without explicit exploration. We use the busy cycle tail bound in \cref{lem:conc-busy-period} to show that the system hits the zero state often enough to give at least $\nicefrac{\theta }{2} \times \sqrt{t}$ samples in the first half of $(0,t]$ (The constant $\theta$ depends on the system parameters). This ensures that the algorithm does not need to explore in the latter half of $(0,t]$.  

\item That the system hits the zero state at least once in the latter half of $(0,t]$
(\cref{lem:cmuhat-hit-zero}).
This can be verified using the busy cycle concentration in
\cref{lem:conc-busy-period}.
\end{enumerate}

Next, we show (in \cref{lem:actual-genie-monotonic}) the following monotonicity
property for the $c\mu$ rule: if two systems with identical parameters,
and initial queue-states satisfying $\bm{Q}(0)\le\bm{Q}^*(0)$ element-wise
follow the $c\mu$ algorithm, then the same ordering of their respective queue-states is maintained in subsequent time-slots, i.e.,
$\bm{Q}(t)\le\bm{Q}^*(t)$ for all $t>0$.

To summarize the argument, we have with polynomially high probability that
(i) the $c\hat{\mu}$ algorithm 
agrees with the $c\mu$ rule while exploiting after
time $\sqrt{t}$ (\cref{lem:correct-order}),
(ii) it only exploits
in the latter half of $(0,t]$ and does not explore (\cref{lem:no-explore}), and
(iii) the system reaches the zero state (which is smaller than any state that the genie system could be in) at least once in the latter half of $(0,t]$
(\cref{lem:cmuhat-hit-zero}).
Thus, the monotonicity property
(in \cref{lem:actual-genie-monotonic} in Appendix~\ref{AppC})
shows that the $c\hat{\mu}$ system always maintains a queue-length not exceeding that of the genie system after it first hits the zero state in the latter half of $(0,t]$. Effectively, at time $t$, the regret is positive only with probability $o(\nicefrac{1}{\mathrm{poly}(t)})$ which gives us the required decay of expected queue-length error in \cref{thm:c-mu-hat}.

The proofs of \cref{thm:c-mu-hat} and
\cref{lem:correct-order,lem:no-explore,lem:cmuhat-hit-zero}
are given in detail in Appendices~\ref{AppC} and \ref{AppD}, respectively.

The degradation of convergence rate of the queue-length error from exponential in a single server system to super-polynomial in a parallel server system can be explained by the addition of explicit exploration in the  $c\hat{\mu}$ algorithm for the latter.
In this situation, we can only show that the  $c\hat{\mu}$ algorithm needs to explore with a probability that vanishes at a polynomial rate.
However, for exponential convergence, one needs to establish that the algorithm deviates from the  $c\mu$ rule with 
a probability that vanishes at an exponential rate. Designing algorithms with
the best achievable convergence rates is an area of future work.

\subsection{Extension to other genie policies.}
We now discuss the scope of generalizing the results in this paper to scheduling policies other than the $c\mu$ algorithm. Consider the bipartite graph with queues and servers as the nodes and the links between them as the edges.
We define a \emph{static priority rule} as a scheduling policy which allocates servers to non-empty queues according to a given priority order for the links. For example, the $c\mu$ rule is a static priority rule where the priority order of the links is given by the descending order of their weights $\{c_i \mu_{i,j}\}$. 
Now, consider genie algorithms that are based on static priority rules, i.e., in every time-slot, the same priority order is used to assign servers to non-empty queues. If the $c\mu$ algorithm is replaced by any static priority genie algorithm, the same proof technique given above can be applied if the monotonicity property in
\cref{lem:actual-genie-monotonic,lem:actual-coupled-monotonic} holds for the corresponding static priority rule.  
This monotonicity property can be proved for any rule with queue priority, i.e., a static priority rule where queues have a specified order of priority and, for each queue, the links are ordered according to their service rates to that queue. Therefore, the regret bound in \cref{thm:c-mu-hat} also holds for algorithms where the exploit rule in \cref{alg:c-mu-hat} is replaced by rules with queue priority.

Moreover, \cref{lem:conc-busy-period} holds for any static priority algorithm,
whereas the region of arrival rates given by Condition~\eqref{eq:feasibility}
depends on the priority rule for a general parallel server system.  
In Appendix~\ref{S6}, we show that exponential tail bounds for busy cycles hold within the entire stability region for a special class of policies that we refer to as \emph{hierarchical rules}.
Thus, for a hierarchical rule that satisfies the monotonicity property,
we can show $O(1)$ regret for \cref{alg:c-mu-hat} (with the $c\mu$ rule replaced by the hierarchical rule)
within the entire stability region.


\begin{APPENDICES}

\section{Stability of hierarchical rules in parallel-server networks.}
\label{S6}


In this section we extend the results of \cref{thm:cmu-instability,T3.2},
and show a special class of rules for which geometric ergodicity holds in the entire stability region.

Consider a queueing network with $U$ classes of customers
and $K$ servers.  
The queues are labeled as $1,\dots, U$ and the servers as $1,\dots, K$.
Set $\sI = \{1,\dots,U\}$ and $\sJ = \{1, \dots, K\}$.  
Each queue can be served by a subset of servers,
and each server can serve a subset of queues.
For each $i \in \sI$, let $\sJ(i) \subset \sJ$
be the subset of servers that can serve queue $i$, and for each
$j \in \sJ$, let $\sI(j) \subset \sI$ be the subset
of queues that can be served by server $j$.
For each $i \in \sI$ and $j \in \sJ$, if queue $i$
can be served by server $j$, we denote $i \sim j$ as an edge in the bipartite
graph formed by the nodes in $\sI$ and $\sJ$;
otherwise, we denote $i \nsim j$.
Let $\sE$ be the collection of all these edges.
Let $\sG = (\sI\cup \sJ, \sE)$ be the
bipartite graph formed by the nodes (vertices) $\sI\cup \sJ$
and the edges $\sE$.
We assume that $\sG$ is connected.

A static priority rule can be identified with a permutation of the edges of the
graph $\sG$, i.e., one--to--one map
$\upsigma\colon\sE\to\lbrace 1,\dotsc,\abs{\sE}\rbrace$ defined by the priority rule --  $\upsigma(e)<\upsigma(e')$ if edge $e$ has higher priority than edge $e'$.
\begin{definition}[Hierarchical Rule]
For a static priority rule $\upsigma$ and for any $i$ and $i'$ with $\sJ(i)\cap \sJ(i')\ne\varnothing$, we say that $i\lessdot i'$ if $\upsigma(i,j)<\upsigma(i',j)$ for all $j\in\sJ(i)\cap \sJ(i')$.
A static priority rule $\upsigma$ is \emph{hierarchical} if
$\lessdot$ defines a partial order on $\sI$, and for any $i$ and $i'$ with $\sJ(i)\cap \sJ(i')\ne\varnothing$, either $i\lessdot i'$ or $i'\lessdot i$.
\end{definition}

It is easy to see that if $\sG$ is a tree, then every static priority rule
is hierarchical. 

In the rest of this section, we study only hierarchical rules.


%

\subsection{Hierarchical decomposition.}\label{S6.2}
Consider a queueing network with graph $\sG$, parameters $\bm\lambda$ and $\bm\mu$,
with $\bm\lambda\in\mathcal{C}(\bm{\mu})$,
under a hierarchical static priority rule $\upsigma$.
We let $\sI^{(1)}=\sI$, $\sJ^{(1)}=\sJ$, $\sE^{(1)}=\sE$,
$\sG^{(1)}= (\sI^{(1)}\cup \sJ^{(1)}, \sE^{(1)})=\sG$,
$\bm\mu^{(1)}=\bm\mu$, and denote by
$\widehat\sI^{(1)}$  the minimal elements of $\sI^{(1)}$ under $\prec_\upsigma$.
The dependence on the arrival rates is suppressed in this notation, since
at each step of the decomposition the arrival rates match the original ones
$\bm\lambda$, while the service rates are modified.

Consider the subgraph with queue nodes $\widehat\sI^{(1)}$ and
server nodes $\widehat\sJ^{(1)}\df\cup_{i\in\widehat\sI^{(1)}} \sJ(i)$.
Since $\widehat\sI^{(1)}$ consists of minimal elements, it follows
that $\sJ(i)\cap\sJ(i')=\varnothing$ if $i,i'\in\widehat\sI^{(1)}$ and $i\ne i'$.
Hence each queue $Q_i$, $i\in\widehat\sI^{(1)}$ forms a Markov process, which
is geometrically ergodic, since $\bm\lambda\in\mathcal{C}(\bm{\mu})$.
Let $\uppi^{(1)}_i$, $i\in\widehat\sI^{(1)}$, denote the stationary distribution
of $Q_i$.

Next, we remove the nodes $\widehat\sJ^{(1)}$ and associated edges from
$\sG^{(1)}$, and denote the resulting graph, which might not be connected, by
$\sG^{(2)}= (\sI^{(2)}\cup \sJ^{(2)}, \sE^{(2)})$.
We let
$\widehat\sI^{(2)}$ denote the minimal elements of $\sI^{(2)}$ under $\prec_\upsigma$.
Removing these nodes and and associated edges from
$\sG^{(2)}$, we obtain a graph $\sG^{(3)}= (\sI^{(3)}\cup \sJ^{(3)}, \sE^{(3)})$,
and so on by induction.
We let $m$ denote the largest integer such that $\sG^{(m)}\ne\varnothing$.

Let $\Breve\sI^{(k)} \df \widehat\sI^{(1)}\cup\dotsb\cup\widehat\sI^{(k)}$,
$k\le m$,
and let $\bm{Q}^{(k)}$ denote the queueing process restricted to $\Breve\sI^{(k)}$.
It is clear that this is Markov.
Provided that it is positive recurrent, we let $\bm{\uppi}^{(k)}$
denote its invariant probability measure.

\subsection{The structure of the transition kernels.}\label{S6.3}
Let $i\in\widehat\sI^{(k+1)}$ for some $k\in\{1,\dotsc,m-1\}$.
It is clear that the transition kernel of $Q_i$ depends on $\bm{q}^{(k)}$,
and thus takes the
form $P_i(q_i' \,|\, q_i, \bm{q}^{(k)})$.
Due to the hierarchical rule, a server $j\in\sJ(i)$ may not
be available to queue $Q_i$
if the queues $\bm{Q}^{(k)}$ have sufficient size.
It is evident then that the transition kernel of $Q_i$ has the following
structure.  There exists a finite partition $\{\sA_{i,\ell}\colon \ell=1,\dotsc,n_i\}$
of $\ZZ_+^{k}$ and associated transition kernels
$\{\sP_{i,\ell}\colon \ell=1,\dotsc,n_i\}$, with each $\sP_{i,\ell}$ corresponding
to a queue with arrival rate $\lambda_i$ and served by a subset of
the servers $\sJ(i)$, such that
\begin{equation}\label{E-kernel0}
P_i(q_i' \,|\, q_i, \bm{q}^{(k)}) \;=\; 
\sum_{\ell=1}^{n_i}\Ind_{\sA_{i,\ell}}(\bm{q}^{(k)})\sP_{i,\ell}(q_i' \,|\, q_i)\,.
\end{equation}

We illustrate this via the following example.
Consider the `W' network in \cref{FigW}.

\begin{figure}[ht]
\centering
\includegraphics[width=0.45\textwidth]{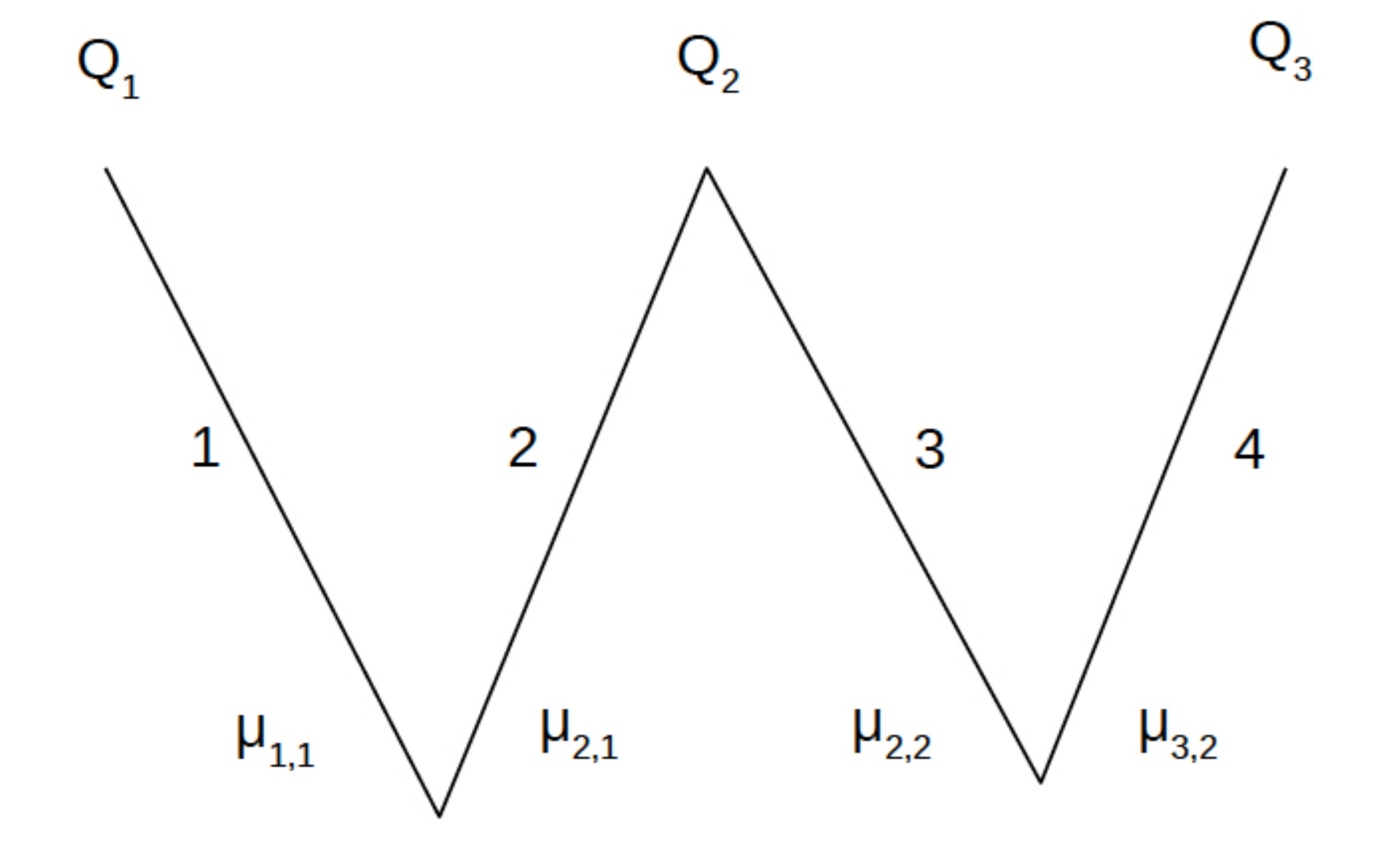}
\caption{Figure to demonstrate the structure of the transition kernels.}\label{FigW}
\end{figure}

It is clear that
\begin{equation}\label{EW-kernA}
P_2(q_2'\,|\,q_2,q_1) \;=\;
\Ind_{\{0\}}(q_1) \sP_{[\lambda_2;\mu_{2,1},\mu_{2,2}]}
+ \Ind_{\{0\}^c}(q_1)\sP_{[\lambda_2;\mu_{2,2}]}\,,
\end{equation}
where we use the notation $\sP_{[\lambda_2;\mu_{2,2}]}$ to denote
the transition kernel of a single-queue, single-server system with parameters
$\lambda_2$ and $\mu_{2,2}$.
Continuing, we also have
\begin{equation}\label{EW-kernB}
P_3\bigl(q_3'\,|\,q_3,(q_1,q_2)\bigr) \;=\;
\Ind_{\sA_{3,1}}(q_1,q_2)\sP_{[\lambda_3,\mu_{3,2}]}+
\Ind_{\sA_{3,2}}(q_1,q_2)\sP_{[\lambda_3]}\,,
\end{equation}
with $\sA_{3,1}=(\ZZ_+\times\{0\})\cup\{(0,1)\}$, and $\sA_{3,2}=\sA_{3,1}^c$.
Here $\sP_{[\lambda_3]}$ corresponds to a transient process,
with arrivals but no service.

Next we discuss the ergodic properties of the `W' network in \cref{FigW}.
Suppose that the arrival rates lie in the capacity region.
Then of course $\lambda_1<\mu_{1,1}$ and $Q_1(t)$ is a geometrically ergodic
Markov chain with stationary distribution $\uppi_1$.
It follows by \cref{EW-kernA} and the proof of \cref{thm:cmu-instability,T3.2} that if
\begin{equation}\label{EW-stabA}
\uppi_1(\{0\})\mu_{2,1} + \mu_{2,2} \,>\, \lambda_2\,,
\end{equation}
then the chain $\bm{Q}^{(2)}=\bigl(Q_1(t),Q_2(t)\bigr)$ is geometrically
ergodic, and if the opposite inequality holds in \cref{EW-stabA}, then
it is transient.
Continuing, assume \cref{EW-stabA}, and let $\bm{\uppi}^{(2)}$ denote the
stationary distribution of $\bm{Q}^{(2)}(t)$.
Applying the same reasoning to \cref{EW-kernB}, it follows that if
\begin{equation}\label{EW-stabB}
\bm{\uppi}^{(2)}(\sA_{3,1})\mu_{3,2}  \,>\, \lambda_3\,,
\end{equation}
then $\bm{Q}^{(3)}=\bigl(Q_1(t),Q_2(t),Q_3(t)\bigr)$
is geometrically
ergodic, otherwise it is not.
Thus, combining the above discussion with \cref{T3.2}, it is clear that
the queueing process $\bigl(Q_1(t),Q_2(t),Q_3(t)\bigr)$ is geometrically ergodic
if and only if \cref{EW-stabA,EW-stabB} hold.

\subsection{The averaged kernel.}\label{S6.4}
Recall the notation introduced in \cref{S6.4}.
Suppose that the queueing process $\bm{Q}^{(k)}$ is geometrically ergodic,
and as introduced earlier, let $\bm{\uppi}^{(k)}$
denote its invariant probability measure.
We define the \emph{averaged kernel} $\overline{P}_i$ of
\cref{E-kernel0} by
\begin{equation*}
\overline{P}_i(q_i' \,|\, q_i) \;=\; 
\sum_{\ell=1}^{n_i}\bm{\uppi}^{(k)}(\sA_{i,\ell})\,\sP_{i,\ell}(q_i' \,|\, q_i)\,.
\end{equation*}
Recall that each kernel $\sP_{i,\ell}$ corresponds to a single-queue system
with arrival rate $\lambda_i$, and service rates $\mu_{i,j}$ for a
subset $\Tilde\sJ_{i,\ell}\subset\sJ(i)$ of the original server nodes
($\Tilde\sJ_{i,\ell}$ might be empty).
For each $j\in\sJ(i)$ define
\begin{equation}\label{ES6.4B}
\widetilde{\sA}_{i,j} \,\df\, \bigcup_{\{\ell\colon j\in \Tilde\sJ_\ell\}} \sA_{i,\ell}\,.
\end{equation}
It is clear  that
\begin{equation*}
\sum_{q_i'\in\ZZ_+} q_i'\sP_{i,\ell}(q_i' \,|\, q_i)  - q_i
\,=\, \lambda_i - \sum_{j\in\Tilde\sJ_{i,\ell}} \mu_{i,j}\qquad
\forall\,q_i\ge\abs{\sJ(i)}\,.
\end{equation*}
A direct computation then shows that
\begin{equation}\label{ES6.4D}
\sum_{q_i'\in\ZZ_+} q_i'\overline{P}_i(q_i' \,|\, q_i)  - q_i
\,=\, \lambda_i - \sum_{j\in\sJ(i)} \bm{\uppi}^{(k)}(\widetilde{\sA}_{i,j})\mu_{i,j}\qquad
\forall\,q_i\ge\abs{\sJ(i)}\,.
\end{equation}
It is evident then that the averaged kernel $\overline{P}_i$
corresponds to a geometrically ergodic chain (transient chain) if the right hand side
of \cref{ES6.4D} is negative (positive).

A proof that is identical to those in \cref{thm:cmu-instability,T3.2} then
asserts the following.
We use the notation introduced in \cref{S6.2,S6.3}, and \cref{ES6.4B}.

\begin{theorem}\label{T6.1}
Consider a queueing network with graph $\sG$, parameters $\bm\lambda$ and $\bm\mu$,
with $\bm\lambda\in\mathcal{C}(\bm{\mu})$,
under a hierarchical rule $\upsigma$.
Suppose that $\bm{Q}^{(k)}$ for some
$k\in\{1,\dotsc,m-1\}$, is geometrically ergodic,
and let $\bm{\uppi}^{(k)}$
denote its invariant probability measure.
Then the chain $\bm{Q}^{(k+1)}$ is geometrically ergodic
if and only if
\begin{equation}\label{ET6.1A}
\sum_{j\in\sJ(i)} \bm{\uppi}^{(k)}(\widetilde{\sA}_{i,j})\mu_{i,j}
\,>\, \lambda_i\qquad \forall\,i\in\widehat\sI^{(k+1)}\,.
\end{equation}
\end{theorem}

The following corollary is immediate from \cref{T6.1}.

\begin{corollary}\label{C6.1}
Consider a queueing network as in \cref{T6.1}.
Then the queueing process is stable iff it is geometrically ergodic,
and this is equivalent to \cref{ET6.1A} for all $k=1,\dotsc,m-1$.
\end{corollary}

\begin{example}
We demonstrate \cref{T6.1} for the hierarchical rule in \cref{FigEx6.1}.
\begin{figure}[ht]
\centering
\includegraphics[width=0.5\textwidth]{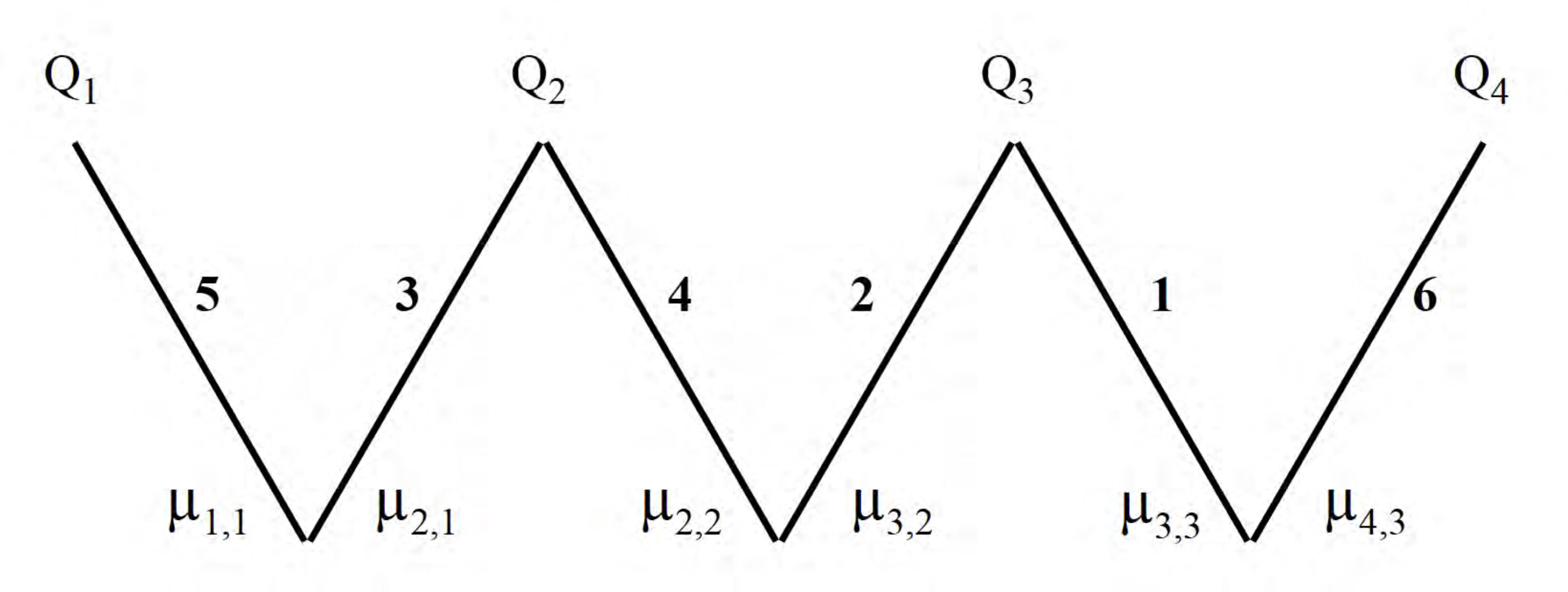}
\caption{A parallel-server network under a hierarchical rule.}\label{FigEx6.1}
\end{figure}
\end{example}
Here $\bm{Q}^{(1)}=Q_3$, $\bm{Q}^{(2)}=(Q_2,Q_3,Q_4)$, and $\bm{Q}^{(3)}=\bm{Q}$.
Necessary and sufficient conditions for stability and geometric ergodicity
are the following.
\begin{align*}
\mu_{3,2}+\mu_{3,3}&\,>\, \lambda_3\,,\\[3pt]
\bm{\uppi}^{(1)}(\{0\})\,\mu_{4,3}&\,>\, \lambda_4\,,\\[3pt]
\mu_{2,1}+\bm{\uppi}^{(1)}(\{0,1\})\,\mu_{2,2}&\,>\, \lambda_2\,,\\[3pt]
\bm{\uppi}^{(2)}\bigl(\{0\}\times\ZZ_+\times\ZZ_+\bigr)\,\mu_{1,1}&\,>\, \lambda_1\,.
\end{align*}

\medskip
\begin{example}
Consider a network such that $1\sim j$ for all $j\in\sJ$ ($\sJ(1) = \sJ$),
and $\sJ(i)$ is a singleton for $i=2,\dotsc, U$.
This class of graphs includes the `N' network and the `W' network,
and we refer to it as a \emph{generalized N network}.
Suppose a hierarchical policy is given such that
$\upsigma(1,j)>\upsigma(i,j)$ for $i\in\sI(j)\setminus\{1\}$, $j\in\sJ$.
It is straightforward to verify that the necessary and sufficient conditions
from \cref{C6.1} are
\begin{equation*}
1- \sum_{i\in\sI(j)\setminus\{1\}}
\frac{\lambda_i}{\mu_{i,j}}\,>\,0\qquad\forall\,j\in\sJ\,,\quad\text{and\ \ }
\sum_{j\in\sJ} \Biggl( 1- \sum_{i\in\sI(j)\setminus\{1\}}
\frac{\lambda_i}{\mu_{i,j}}\Biggr)\mu_{1,j} \,>\,\lambda_1\,.
\end{equation*}
Thus the queueing process is geometrically ergodic for all
$\bm\lambda\in\mathcal{C}(\bm{\mu})$.
\end{example}

\medskip
\begin{example}
This is an example of a queueing network with
$\bm\lambda\in\mathcal{C}(\bm{\mu})$
that is not stabilizable under
any static priority policy.
Consider a a $2 \times 3$ `M' network.
For $\epsilon>0$ a scaling parameter, we choose $\bm\lambda$ and $\bm\mu$ such
that $\lambda_1 = \mu_{1,1} + \epsilon$,
$\lambda_2 = \mu_{2,3} + \epsilon$, and $\mu_{1,2}=\mu_{2,2}=3\epsilon$.
Thus
\begin{equation*}
\frac{\lambda_1-\mu_{1,1}}{\mu_{1,2}} =\dfrac{1}{3}<\frac{2}{3}=
\frac{\mu_{2,2}+\mu_{2,3} -\lambda_2}{\mu_{2,2}}\,,
\end{equation*}
which shows that $\bm\lambda\in\mathcal{C}(\bm{\mu})$.

Now, let $\uppi_1$ ($\uppi_2$) denote the stationary distribution of $Q_1$ ($Q_2$) under the static priority policy that gives higher priority to Queue $1$ (Queue $2$) in assigning Server $2$. It can be shown that we can select $\epsilon$ sufficiently small so that
$$\min\{\uppi_1(\{0,1\}), \uppi_2(\{0,1\})\} < \frac{1}{3}.$$
For such an $\epsilon$, we have
\begin{equation*}
\uppi_2(\{0,1\}) \mu_{1,2} + \mu_{1,1} 
< \lambda_1\,,\quad\text{and\ \ }
\uppi_1(\{0,1\}) \mu_{2,2} + \mu_{2,3} < \lambda_2\,,
\end{equation*}
which shows that no static priority policy is stable.
\end{example}




\section{Proofs of \texorpdfstring{\cref{thm:cmu-instability,T3.2}}{}.}\label{AppA}
\proof{Proof of \cref{thm:cmu-instability}.}

Consider a $2 \times 2$ system with service rates $\bm{\mu}$ such that
$\mu_{2,1} < \mu_{1,1},\; \mu_{2,2} < \mu_{1,2},$ and arrival rates
$\bm\lambda\in\mathscr{C}(\bm{\mu})$ under the $c\mu$ rule.
Let the arrivals, instantaneous service rates and queue-lengths of the $2 \times 2$ system at time $t$ be denoted by $\mathbf{A}(t)$, $\mathbf{R}(t)$ and $\mathbf{Q}(t)$ respectively. Without loss of generality, we assume that $\mathbf{Q}(1) = \mathbf{0}$.

Now, consider the Markov chain $\bigl\lbrace Q_1(t) \bigr\rbrace_{t \ge 0}$. We show that this Markov chain is geometrically ergodic, i.e., there exists a function
$V\colon \ZZ_+ \rightarrow [1, \infty]$, a finite set $\sB$,
and constants $\gamma < 1$, $b < \infty$ such that
\begin{equation}
\label{eq:drift-geom-ergodic}
\EE\left[ V(Q_1(t+1)) \given Q_1(t) \right] \le \gamma V(Q_1(t)) + b \ind{Q_1(t) \in \sB}.
\end{equation}
Since $\lambda_1 < \mu_{1,1} + \mu_{1,2}$, we have
\begin{align*}
\EE\left[ e^ { \left( R_{1,1}(t) + R_{1,2}(t) \right)} \right] & = (1 - \mu_{1,1})(1 - \mu_{1,2}) + \left( (1 - \mu_{1,1}) \mu_{1,2} + (1 - \mu_{1,2}) \mu_{1,1} \right) e + \mu_{1,1} \mu_{1,2} e^ {2}	\\
& = 1 + \left( \mu_{1,1} + \mu_{1,2} \right) \left( e - 1 \right) + \mu_{1,1} \mu_{1,2} \left( e - 1 \right)^2	\\
& > 1 + \lambda_1 \left( e - 1 \right)	\\
& = \EE\left[ e^ { A_1(t)} \right].
\end{align*}
Therefore, the drift condition~\eqref{eq:drift-geom-ergodic} is satisfied
with
\begin{equation*}
V(x) = \E^x, \quad \sB = \{0, 1\},\quad b = \E^2,\quad\text{and\ \ }
\gamma = \EE\left[ e^ {  A_1(t)} \right]
\Bigl(\EE\Bigl[ e^ { \left( R_{1,1}(t) + R_{1,2}(t) \right)} \Bigr]\Bigr)^{-1}.
\end{equation*}

Let the function $f\colon \ZZ_+ \to [0,2]$ be defined as
\begin{equation*}f(q_1) \df \ind{q_1 = 0} \mu_{2,1} + \ind{q_1 \in \{0,1\}} \mu_{2,2}.\end{equation*}
Since $\bigl\lbrace Q_1(t) \bigr\rbrace_{t \ge 0}$  is geometrically ergodic, from \cite[Theorem 0.2]{dedecker2015subgaussian}, there exists a constant $b_0$ such that for any $\epsilon > 0$ and  $t \in \NN$,
\begin{equation}
\label{eq:geom-ergodic-conc}
\PP\left[ \sum_{l=1}^{t} f(Q_1(l)) > \EE\left[ \sum_{l=1}^{t}  f(Q_1(l)) \right] + 2 \epsilon t \right] \le e^ { -2b_0 \epsilon^2 t}.
\end{equation}
Moreover, since $Q_1(1) \in \sB$, from \cite[Lemma 0.7]{dedecker2015subgaussian}, we can conclude that there exists a constant $M < \infty$ such that for any $t \in \NN$,
\begin{equation}
\label{eq:geom-ergodic-mean-conv}
\EE\left[ \sum_{l=1}^{t}  f(Q_1(l))  \right] \le
\bigl( \uppi_1(0) \mu_{2,1}  + \uppi_1(\{0,1\}) \mu_{2,2} \bigr) t + M.
\end{equation}
Now, let 
\begin{equation*}
b_2 \df \frac{1}{4} \bigl(\lambda_2 - \uppi_1(0) \mu_{2,1} - \uppi_1(\{0,1\})
\mu_{2,2} \bigr) ,
\end{equation*}
and fix positive constants $\epsilon_a$, $\epsilon_b$, $\epsilon_c$ such that
\begin{equation}
\label{eq:const-ineq}
\epsilon_a + 2 \epsilon_b + 2 \epsilon_c \le 2b_2.
\end{equation}
Given the events
\begin{equation}
\label{eq:arrival-conc}
\sum_{l=1}^{t} A_2(l) \ge (\lambda_2 - \epsilon_a) t,
\end{equation}
\begin{equation}
\label{eq:service-conc}
\sum_{l=1}^{t} \left( \ind{Q_1(l) = 0} R_{2,1}(l) + \ind{Q_1(l) \in \{0,1\}} R_{2,2}(l) \right)	 \le \sum_{l=1}^{t}  f(Q_1(l))  + 2 \epsilon_b t,
\end{equation}
and
\begin{equation}
\label{eq:q1-small-conc}
\sum_{l=1}^{t} f(Q_1(l)) > \EE\left[ \sum_{l=1}^{t}  f(Q_1(l)) \right] + 2 \epsilon_c t,
\end{equation}
 we have
\begin{align}
Q_2(t) & \ge \sum_{l=1}^{t} \Bigl( A_2(l) - \ind{Q_1(l) = 0} R_{2,1}(l) - \ind{Q_1(l) \in \{0,1\}} R_{2,2}(l) \Bigr)	\nonumber	\\
& \ge (\lambda_2 - \epsilon_a) t - \sum_{l=1}^{t}  f(Q_1(l)) - 2 \epsilon_b t	\label{eq:q2-lb1}	\\
& \ge (\lambda_2 - \epsilon_a - 2 \epsilon_b) t - \EE\left[ \sum_{l=1}^{t}  f(Q_1(l)) \right] - 2 \epsilon_c t	\label{eq:q2-lb2}	\\
& \ge (\lambda_2 - \epsilon_a - 2 \epsilon_b - 2 \epsilon_c) t -
\Bigl(\left( \uppi_1(0) \mu_{2,1}  + \uppi_1(\{0,1\}) \mu_{2,2} \right) t + M \Bigr)	\label{eq:q2-lb3}	\\
& \ge \Bigl( \lambda_2 - \left( \uppi_1(0) \mu_{2,1}  + \uppi_1(\{0,1\}) \mu_{2,2} \right) - \left( \epsilon_a + 2 \epsilon_b + 2 \epsilon_c \right) \Bigr) t - M	\nonumber	\\[3pt]
& \ge 2b_2 t - M,	\label{eq:q2-lb4}
\end{align}
where lower bound \eqref{eq:q2-lb1} follows from \eqref{eq:arrival-conc} and \eqref{eq:service-conc}, \eqref{eq:q2-lb2} from \eqref{eq:q1-small-conc}, \eqref{eq:q2-lb3} from \eqref{eq:geom-ergodic-mean-conv}, and  \eqref{eq:q2-lb4} from the definition of $b_2$ and \eqref{eq:const-ineq}.

Using the Azuma--Hoeffding bound for bounded martingales, we can obtain the inequalities
\begin{equation*}
\PP\left[ \eqref{eq:arrival-conc}\text{ is false } \right] \le \exp(-2 \epsilon_a^2 t),
\end{equation*}
\begin{equation*}
\PP\left[ \eqref{eq:service-conc}\text{ is false } \right] \le \exp(-2 \epsilon_b^2 t).
\end{equation*}
These along with \eqref{eq:geom-ergodic-conc} give
\begin{equation*}
\PP\left[ Q_2(t) < 2b_2 t - M \right] \le
\exp(-2 \epsilon_a^2 t) + \exp(-2 \epsilon_b^2 t) + \exp(-2b_0 \epsilon_c^2 t).
\end{equation*}
Therefore, for $b_1 \df \min\left( \epsilon_a^2, \epsilon_b^2, b_0 \epsilon_c^2 \right)$
and 
\begin{equation*}
t_0 \df \min \left\lbrace l \ge \frac{M}{b_2}: \exp(-2 \epsilon_a^2 t)
+ \exp(-2 \epsilon_b^2 t) + \exp(-2b_0 \epsilon_c^2 t) \le
\exp\left( -b_1 t \right) \; \forall t \ge l \right\rbrace,
\end{equation*}
we have the required result, i.e.,
\begin{equation*}
\PP\left[ Q_2(t) < b_2 t \right] \le \exp(-b_1 t) \quad \forall t \ge t_0\,.\qquad\Halmos
\end{equation*}
\endproof

\proof{Proof of \cref{T3.2}.}
Let $\Lyap_1(q_1)\df \E^{\delta_1 q_1}$
The process $Q_1(t)$ is time-homogeneous Markov and satisfies the
drift inequality
\begin{equation}\label{PT3.2A}
\EE\bigl[\Lyap_1\bigl(Q_1(t+1)\bigr)\mid Q_1(t)=q_1\bigr] - \Lyap_1(q_1)
\;\le\;  \Tilde\kappa_0\Ind_{\{q_1=0,1\}} - \Tilde\kappa_1 \Lyap_1(q_1)\,, 
\end{equation}
for some positive constants $\Tilde\kappa_0$ and $\Tilde\kappa_1$ which depend
only on $\delta_1$ and the parameters, for all $\delta_1>0$ sufficiently small.
Let $\widetilde{P}_t(q_1,\cdot\,)$ denote the $t$-step transition
probability of $Q_1(t)$.
It is well known (see \cite{Meyn-Tweedie}) that \cref{PT3.2A} implies that
$Q_1$ is geometrically ergodic, and with $\uppi_1$ denoting its
stationary distribution we have
\begin{equation}\label{PT3.2B}
\bnorm{\widetilde{P}_t(q_1,\,\cdot\,)-\uppi_1(\,\cdot\,)}_{\mathsf{TV}} \;\le\;
\widetilde{C}_1\,\Lyap_1(q_1)\, \rho_1^t\qquad\forall\,t\ge0\,.
\end{equation}
for some positive constants $\widetilde{C}_1$ and $\rho_1\in(0,1)$, which depend
on $\delta_1$.

We first show that the chain $\bm{Q}(t)=\bigl(Q_1(t),Q_2(t)\bigr)$
is positive recurrent under the $c\mu$ rule.
We calculate $F(q_1,q_2)\df
\EE\bigl[Q_2(t+1)\mid Q_1(t)=q_1, Q_2(t)=q_2\bigr] - q_2$
for different values of $q_1$ and $q_2$.
Recall that
\begin{equation*}
c_2 \mu_{2,1} < c_1 \mu_{1,1},\quad c_2 \mu_{2,2} < c_1 \mu_{1,2},
\quad \mu_{1,2} < \mu_{1,1}.
\end{equation*}
We also  assume, without loss of
generality, that $\mu_{2,2}>\mu_{2,1}$.
Thus we have
\begin{equation*}
\begin{aligned}
F(q_1,q_2)  &\,=\, \lambda_2-\mu_{2,1}-\mu_{2,2}\,, &&\text{if\ }q_2>1\,,\ q_1=0\,,\\[5pt]
F(q_1,q_2) &\,=\, \lambda_2-\mu_{2,2}\,, &&\text{if\ }q_2=1\,,\ q_1=0\,,\\[5pt]
F(q_1,q_2) &\,=\,\lambda_2-\mu_{2,2}\,, &&\text{if\ } q_2>0\,,\ q_1=1\,,\\[5pt]
F(q_1,q_2) &\,=\, \lambda_2\,,&&\text{if\ } q_1\in\{0,1\}^c\,,\\[5pt]
F(q_1,q_2) &\,=\,\lambda_2\,, &&\text{if\ } q_2=0\,.
\end{aligned}
\end{equation*}
So we can write
\begin{equation*}
F(q_1,q_2)\,=\, \lambda_2 - \Ind_{\{0\}}(q_1)\mu_{2,1}- \Ind_{\{0,1\}}(q_1)\mu_{2,2}
+\Tilde{F}(q_1,q_2)\,,
\end{equation*}
with
\begin{equation*}
\Tilde{F}(q_1,q_2)\,=\, \mu_{2,1}\Ind_{\{0\}}(q_1)\Ind_{\{0,1\}}(q_2)
+ \mu_{2,2}\Ind_{\{0,1\}}(q_1)\Ind_{\{0\}}(q_2)\,.
\end{equation*}
Let $\PP_q$ denote the probability measure on the canonical space
of the chain $\bm{Q}(t)$ starting from $(Q_1(0),Q_2(0))=q=(q_1,q_2)$,
and $\EE_q$ the corresponding expectation operator.
Defining $\Lambda\df\lambda_2 - \uppi_1(\{0\})\mu_{2,1}- \uppi_1(\{0,1\})\mu_{2,2}$,
and using \cref{PT3.2B}, we obtain
\begin{align}\label{PT3.2C}
\EE_q\bigl[F\bigl(Q_1(t),Q_2(t)\bigr)\bigr]
&\,=\, \lambda_2 - \mu_{2,1}\PP_q\bigl(Q_1(t)=0\bigr)
- \mu_{2,2}\PP_q\bigl(Q_1(t)\in\{0,1\}\bigr)
+\EE_q\bigl[\Tilde{F}\bigl(Q_1(t),Q_2(t)\bigr)\bigr]
\nonumber\\[3pt]
&\,=\,\Lambda - \bigl[\PP_q\bigl(Q_1(t)=0\bigr)-\uppi_1(\{0\})\bigr]\mu_{2,1}
-\bigl[\PP_q\bigl(Q_1(t)\in\{0,1\}\bigr)-\uppi_1(\{0,1\})\bigr]\mu_{2,2}\nonumber\\[3pt]
&\mspace{300mu}
+\EE_q\bigl[\Tilde{F}\bigl(Q_1(t),Q_2(t)\bigr)\bigr]\nonumber\\[3pt]
&\,\le\,\Lambda
+\widetilde{C}_1\Lyap_1(q_1) \bigl(\mu_{2,1}+\mu_{2,2}\bigr)\rho_1^t
+\EE_q\bigl[\Tilde{F}\bigl(Q_1(t),Q_2(t)\bigr)\bigr]\,.
\end{align}
Thus, using \cref{PT3.2C} in a telescoping series, we have
\begin{equation*}
\EE_q\bigl[Q_2(T)\bigr] - q_2 \,\le\,\Lambda T
+\frac{\widetilde{C}_1}{1-\rho_1}\bigl(\mu_{2,1}+\mu_{2,2}\bigr)\Lyap_1(q_1)
+\sum_{t=0}^{T-1}\EE_q\bigl[\Tilde{F}\bigl(Q_1(t),Q_2(t)\bigr)\bigr]\,.
\end{equation*}
Dividing by $T$ and letting $T\to\infty$, we obtain
\begin{equation*}
\liminf_{T\to\infty}\;\frac{1}{T}\,
\sum_{t=0}^{T-1}\PP_q\bigl(Q_1(t)\in\{0,1\},Q_2(t)\in\{0,1\}\bigr)
\,\ge\, - \frac{\Lambda}{\mu_{2,1}+\mu_{2,2}}\,,
\end{equation*}
where we use the fact that
$\Tilde{F}(q_1,q_2)\,\le\, (\mu_{2,1}+\mu_{2,2})\Ind_{\{0,1\}}(q_1)\Ind_{\{0,1\}}(q_2)$.
Suppose that $\Lambda<0$.
Since the chain $\bm{Q}(t)$ is irreducible and aperiodic, this shows
that it has a unique invariant probability measure $\uppi$, thus
establishing positive recurrence.

We next strengthen this to geometric ergodicity.
Let $V(q_1,q_2)$ denote the mean hitting time to $(0,0)$ starting from
$(q_1,q_2)$.  We write this as
$V(q_1,q_2)=\EE_q \bigl[\uptau(0,0)\bigr]$.
This is finite by positive recurrence.
It is clear by  the strong Markov property that
$V(q_1,q_2+1)\le\EE_{(q_1,q_2+1)}\bigl[\uptau(0,\{0,1\})\bigr] +
\EE_{(0,1)}\bigl[\uptau(0,0)\bigr]$.
Also by the monotonicity property, we have
$\EE_q \bigl[\uptau(0,0)\bigr]\ge\EE_{(q_1,q_2+1)}\bigl[\uptau(0,\{0,1\})\bigr]$.
Thus $V(q_1,q_2+1)-V(q_1,q_2)$ is bounded uniformly in $(q_1,q_2)$.
Another way of arguing is as follows.
Consider two chains, the first starting at $(q_1,q_2)$, and the
second starting at $(q_1,q_2+1)$ along a given sample path (that is a given
realization of the Bernoulli variables).
Then at the time the first chain hits $(0,0)$ the second chain is
at either $(0,1)$ or $(0,0)$.
It is evident then by the preceding argument that the number $M$ defined by
\begin{equation}\label{E-M}
M\,\df\, \max\,\bigl\{ V(q_1+i,q_2+j)-V(q_1,q_2)\,\colon i,j\in\{0,1\}\,,\;
(q_1,q_2)\in\ZZ_+^2\bigr\}
\end{equation}
is finite.

Given that $M$ in \cref{E-M} is finite, the theorem is a direct consequence of
\cite[Theorem~2.1]{Spieksma-94}, together with the fact that
$V(\bm{q})\ge c \norm{\bm{q}}_1$ for some constant $c>0$.
This completes the proof.

Next, we consider the case $\Lambda\ge0$.
We argue by contradiction.
Suppose that $\bm{Q}(t)$ is stable.
Note that \cref{PT3.2C} can also be written in the form
\begin{equation}\label{PC3.1A}
\EE_q\bigl[F\bigl(Q_1(t),Q_2(t)\bigr)\bigr]\,\ge\,\Lambda
-\widetilde{C}_1\Lyap_1(q_1) \bigl(\mu_{2,1}+\mu_{2,2}\bigr)\rho_1^t
+\EE_q\bigl[\Tilde{F}\bigl(Q_1(t),Q_2(t)\bigr)\bigr]\,.
\end{equation}
Using \cref{PC3.1A} in a telescoping series, together
with  the fact that
$\Tilde{F}(q_1,q_2)\,\ge\, (\mu_{2,1}+\mu_{2,2}) \Ind_{\{(0,0)\})}(q_1,q_2)$,
we obtain
\begin{equation*}
\liminf_{T\to\infty}\;\frac{1}{T}\,
\sum_{t=0}^{T-1}\PP_q\bigl(Q_1(t)=0,Q_2(t)=0\bigr)
\,\le\, \liminf_{T\to\infty}\;\frac{1}{T}\,\EE_q\bigl[Q_2(T)\bigr]\,=\,0
\end{equation*}
where the equality in the above display follows by the
hypothesis that $\bm{Q}(t)$ is stable.
Thus $\bm{Q}(t)$ is not positive recurrent, and this contradicts the
hypothesis that it is stable.
\Halmos\endproof

\section{Proof of Lemma~\ref{lem:conc-busy-period}.}\label{AppB}

Let $\bm{\alpha} \in \mathcal{P}_U$ be such that \cref{eq:feasibility} is satisfied.
We first note that the set of all possible queue-server assignments made by the $c\mu$ rule at time $l$ when $\norm{\bm{Q}(l)}_1 \ge K$ is the same as the set of all possible assignments when $\norm{\bm{Q}^*(l)}_1 = K$. Therefore,
\begin{equation*}
\min_{l \ge K} \min_{\bm{q} \in \mathscr{Q}_l} (\bm{R}(\bm{q}) \cdot \bm{\alpha}) = \min_{\bm{q} \in \mathscr{Q}_K} (\bm{R}(\bm{q}) \cdot \bm{\alpha}) > \bm{\lambda} \cdot \bm{\alpha}.
\end{equation*}
Let $\epsilon_0 \df \bm{\lambda} \cdot \bm{\alpha} - \min_{\bm{q} \in \mathscr{Q}_K} (\bm{R}(\bm{q}) \cdot \bm{\alpha})$. Now, consider the process $\{Y(l) \df \bm{Q}(l) \cdot \bm{\alpha}\}$. We use the drift conditions in \cite{hajek1982hitting} to obtain first hitting time bounds. Note that if $Y(l) \ge K \norm{\bm{\alpha}}_{\infty}$,
then $\norm{\bm{Q}(l)}_1 \ge K$ and therefore, the one-step drift satisfies
\begin{equation*}
\EE\bigl[ Y(l+1) - Y(l) \given \bm{Q}(l), Y(l) > K \norm{\bm{\alpha}}_{\infty} \bigr] \le \bm{\lambda} \cdot \bm{\alpha} - \bm{R}(\bm{Q}(l)) \cdot \bm{\alpha}	 \le  -\epsilon_0.
\end{equation*}
In addition, we also have that $\card{Y(l+1) - Y(l)} \le K \norm{\bm{\alpha}}_{\infty}$. Therefore, from \cite[Theorem 2.3]{hajek1982hitting}, there exist constants $\eta$ and $0 < \rho < 1$ such that, for any  $l > \sqrt{t}$,  the first hitting time of the process $Y$ to the set $\{y\le K \norm{\bm{\alpha}}_{\infty}\}$, denoted as $\uptau_{K}$, satisfies
\begin{equation}
\label{eq:hajek-hitting-time}
\EE\left[ s^{\uptau_{K}} \given \bm{Q}(0) \right]  \le e^{\eta (Y(0) -  K \norm{\bm{\alpha}}_{\infty})} \frac{s-1}{1 - \rho s} + 1 \qquad \forall s\in(1,\rho).
\end{equation}
We use this result to show exponential tail bounds for busy periods. Let
\begin{equation*}
\mathscr{Y}_K  \df \{\bm{q} \neq \bm{0}\,\colon \bm{q} \cdot \bm{\alpha} \le K \norm{\bm{\alpha}}_{\infty}\}.
\end{equation*}
The Markov process $\{\bm{Q}(l)\}_{l > 0}$ is irreducible. Therefore, there exists an integer $m > 0$ such that
\begin{equation*}
p_{\epsilon} \df \min_{\bm{q} \in \mathscr{Y}_K} \PP\left[ \uptau_{\bm{0}} \le m \given \bm{Q}(0) = \bm{q} \right] > 0.
\end{equation*} 
Note that $x^m \left( e^{\eta m } \left( \frac{x - 1}{1 - \rho x} \right) + 1 \right)$ is a continuous increasing function of $x$ and takes the value $1$ at $x = 1$. Therefore, we can find an $s \in (1, \rho^{-1})$ such that 
\begin{equation*}
s^m \left( e^{\eta m } \left( \frac{s - 1}{1 - \rho s} \right) + 1 \right) < \frac{1}{1 - p_{\epsilon}}.
\end{equation*}
Let
\begin{equation*}
M \df \max_{\bm{q} \in \mathscr{Y}_K} \EE\left[ s^{\uptau_{\bm{0}}} \given \bm{Q}(0) = \bm{q} \right],\quad\text{and\ \ }
\bm{q}^*  \in \argmax_{\bm{q} \in \mathscr{Y}_K} \EE\left[ s^{\uptau_{\bm{0}}} \given \bm{Q}(0) = \bm{q} \right].
\end{equation*}
Also, let
\begin{equation*}
\mathscr{Y}_{K;m}  \df \{\bm{q} \neq \bm{0} : 0 < \bm{q} \cdot \bm{\alpha} - K \norm{\bm{\alpha}}_{\infty} \le m  \}.
\end{equation*}
We can obtain an upper bound for $M$ as follows.
\begin{align*}
M & = \EE\left[ s^{\uptau_{\bm{0}}} \given \bm{Q}(0) = \bm{q}^* \right]\\[3pt]
& \le p_{\epsilon} s^m + \sum_{\bm{q} \in \mathscr{Y}_K} \PP\left[ \bm{Q}(m) = \bm{q}, \uptau_{\bm{0}} > m \given[\big] \bm{Q}(0) = \bm{q}^* \right] s^m M	\\[3pt]
&\mspace{50mu} + \sum_{\bm{q} \in \mathscr{Y}_{K + mU}} \PP\left[ \bm{Q}(m) = \bm{q}, \uptau_{\bm{0}} > m \given[\big] \bm{Q}(0) = \bm{q}^* \right]
 s^m \EE\left[ s^{\uptau_{K}} \given[\big] \bm{Q}(0) = \bm{q} \right] M	\\[3pt]
& \le p_{\epsilon} s^m + (1 - p_{\epsilon}) s^m M \left( e^{\eta m \norm{\bm{\alpha}}_{1}} \frac{s-1}{1 - \rho s} + 1 \right),
\end{align*}
where the last inequality follows from \cref{eq:hajek-hitting-time}. This gives us
\begin{equation*}
M \le \frac{p_{\epsilon} s^m}{1 - (1 - p_{\epsilon}) s^m \left( e^{\eta m \norm{\bm{\alpha}}_{1}} \frac{s-1}{1 - \rho s} + 1 \right)} < \infty.
\end{equation*}
For any $\bm{q} \in \mathscr{Y}_K^c$, we have
\begin{align*}
\EE\left[ s^{\uptau_{\bm{0}}} \given[\big] \bm{Q}(0) = \bm{q} \right] & \le \EE\left[ s^{\uptau_{K}} \given[\big] \bm{Q}(0) = \bm{q} \right] M	\\[5pt]
 & \le M \left( e^{\eta (\bm{q} \cdot \bm{\alpha} -  K \norm{\bm{\alpha}}_{\infty})} \frac{s-1}{1 - \rho s} + 1 \right).
\end{align*}
Using the Chernoff bound, for any $r \in \NN$, we get
\begin{align*}
\PP\left[ \uptau_{\bm{0}} > r \given[\big] \bm{Q}(0) = \bm{q} \right] & \le s^{-r} \EE\left[ s^{\uptau_{\bm{0}}} \given[\big] \bm{Q}(0) = \bm{q} \right]	\\[5pt]
 & \le s^{-r} M \left( e^{\eta (\bm{q} \cdot \bm{\alpha} -  K \norm{\bm{\alpha}}_{\infty})} \frac{s-1}{1 - \rho s} + 1 \right)	\\
 & \le \frac{1}{t^{\kappa}}
\end{align*}
if
\begin{equation*}
r \ge \frac{\kappa \log t}{\log s} + \frac{\eta}{\log s} \bm{q} \cdot \bm{\alpha} + \frac{1}{\log s} \left( \log (2M) + \left( \log \left( \frac{s-1}{1 - \rho s} \right) \right)^+ - \eta K \norm{\bm{\alpha}}_{\infty} \right).\end{equation*}
This gives us the desired result by choosing $C_{\ref{const:3}} = \frac{1}{\log s}$,  $C_{\ref{const:4}} = \frac{\eta}{\log s}$ and
\begin{equation*}
C_{\ref{const:5}} = \frac{1}{\log s} \left( \log (2M) + \left( \log \left( \frac{s-1}{1 - \rho s} \right) \right)^+ - \eta K \norm{\bm{\alpha}}_{\infty} \right)\,.\qquad\Halmos
\end{equation*}

\section{Proof of \texorpdfstring{\cref{thm:c-mu-hat}}{}.}\label{AppC}


In the proof of  \cref{thm:c-mu-hat} we use probability estimates
of the following events.
\begin{enumerate}
\item $\mathscr{E}_{\ref{ev:accurate-est}}$:
after time $\sqrt{t}$ the
$c\Hat\mu$ algorithm agrees with the $c\mu$ rule.
\item \event{ev:no-explore}$\mathscr{E}_{\ref{ev:no-explore}}$: $\varepsilon(l) = 0$ for all $l \in (\nicefrac{t}{2}, t]$, where $\varepsilon$ is as in
\cref{alg:c-mu-hat}.
\item \event{ev:cmuhat-hit-zero}$\mathscr{E}_{\ref{ev:cmuhat-hit-zero}}$: $\bm{Q}(l) = \bm{0}$ for some $l \in (\nicefrac{t}{2}, t]$.
\item \event{ev:cmu-hit-zero}$\mathscr{E}_{\ref{ev:cmu-hit-zero}}$: $\bm{Q}^*(l) = \bm{0}$ for some $l \in (\nicefrac{t}{2}, t]$.
\end{enumerate}
The following lemmas show that the probability of the complement of these events decays super polynomially with $t$. Specifically, for any $k \in \NN$, we have the following.

\begin{lemma}
\label{lem:correct-order}
$\PP\left[ \mathscr{E}_{\ref{ev:accurate-est}}^c \right] = o\left( \frac{1}{t^{k+1}} \right)$.
\end{lemma}

\begin{lemma}
\label{lem:no-explore}
$\PP\left[ \mathscr{E}_{\ref{ev:no-explore}}^c \cap \mathscr{E}_{\ref{ev:accurate-est}} \right] = o\left( \frac{1}{t^{k+1}} \right)$.
\end{lemma}

\begin{lemma}
\label{lem:cmuhat-hit-zero}
$\PP\left[ \mathscr{E}_{\ref{ev:cmuhat-hit-zero}}^c \cap \mathscr{E}_{\ref{ev:accurate-est}}\right] = o\left( \frac{1}{t^{k+1}} \right)$.
\end{lemma}

\begin{lemma}
\label{lem:cmu-hit-zero}
$\PP\left[ \mathscr{E}_{\ref{ev:cmu-hit-zero}}^c \right] = o\left( \frac{1}{t^{k+1}} \right)$.
\end{lemma}

\begin{remark}
In both Appendix~\ref{AppC} and Appendix~\ref{AppD}, we use order notation to denote asymptotics with increasing $t$ (and not with increasing $T$ as in the main section). The constant factors in these asymptotic guarantees depend on $k$
 and the parameters of the system.
\end{remark}
 Proofs of \cref{lem:correct-order,lem:no-explore,lem:cmuhat-hit-zero} are given in Appendix~\ref{AppD}. We skip the proof of \cref{lem:cmu-hit-zero} as it is similar to that of \cref{lem:cmuhat-hit-zero}.

Now, recall that the service offered by each of the links is i.i.d.\ across time with mean given by the rate matrix $\bm{\mu}$. In order to prove \cref{thm:c-mu-hat}, we use the following alternate construction of the service process for the $c\hat{\mu}$ and $c\mu$ systems, which gives the same regret as the original service process.
For each $i \in [U]$, let $\{U_{i}(s)\}_{s>0}$ be a sequence of independent
$\text{Unif}(0, 1)$ random variables. Also, define the sequence of random variables $\{Z_{i}(s)\}_{s \ge 0}$ as follows: 
\begin{equation*}
Z_{i}(s)  = Z_{i}(s-1) + \max\{Q_i(s), Q^*_i(s)\}\quad\forall s \in \NN\,,
\qquad Z_{i}(0)= 0\,.
\end{equation*}
 
Now, consider any time slot $l \in \NN$. For both the systems, let us index the jobs at time $l$ in the each of the queues in the order of their arrival, i.e., smaller index for earlier arrivals. Since the service discipline is FCFS, this is also the order of their service. If a job with index $n$ in queue $i$ is assigned server $j$, then it gets service equal to $\ind{U_i(Z_{i}(l-1) + n) > 1-\mu_{i,j}}$. To verify that this construction generates service realizations that are independent across time and with the correct mean rate, note that for any $n>0$,
\begin{equation*}
\EE\left[ \ind{U_i(Z_{i}(l-1) + n) > 1-\mu_{i,j}} \given \mathcal{F}_{l-1} \right]
= \mu_{i,j},
\end{equation*}
where $\mathcal{F}_{l-1}$ is the history at time $l-1$, i.e.,
the $\sigma$-algebra of all the random variables generated until time $l-1$.
For this modification of the service process,
we have $\bm{Q}(t) = \bm{Q}^*(t)$ with high probability,
as asserted by the following lemma.

\begin{lemma}\label{lem:actual-genie-monotonic}
If $t \ge 4$, then
$\PP\bigl[\bm{Q}(t) = \bm{Q}^*(t) \,|\,
\mathscr{E}_{\ref{ev:accurate-est}} \cap \mathscr{E}_{\ref{ev:no-explore}}
\cap \mathscr{E}_{\ref{ev:cmuhat-hit-zero}} \cap \mathscr{E}_{\ref{ev:cmu-hit-zero}}\bigr]=1$ a.s.
\end{lemma}

\proof{Proof.}
Consider the two systems ($c\hat{\mu}$ and $c\mu$) at any time $l \in \NN$. 
Now, note that, if $\bm{Q}(l) \le \bm{Q}^*(l)$, and if at time $l$ the $c\hat{\mu}$ algorithm agrees with the $c\mu$ rule, then the following monotonicity property is satisfied:
For every assignment made in the genie system, the corresponding job (with the same index) in the $c\hat{\mu}$ system, if it exists, is assigned
to a server with higher or equal success probability.
For the modified service process given above, this monotonicity property
implies that, for any job that is successfully served in the $c\mu$ system,
the corresponding job in the $c\hat{\mu}$ system, if it exists, is also
successfully served.
Therefore $\bm{Q}(l+1) \le \bm{Q}^*(l+1)$. 

Now, given $\mathscr{E}_{\ref{ev:accurate-est}} \cap \mathscr{E}_{\ref{ev:no-explore}} \cap \mathscr{E}_{\ref{ev:cmuhat-hit-zero}}$, we have $\bm{Q}(l) = \mathbf{0} \le \bm{Q}^*(l)$ for some
$l \in (\nicefrac{t}{2}, t]$. Since the $c\hat{\mu}$ algorithm follows the $c\mu$
rule with correct ordering in the interval $(\nicefrac{t}{2}, t]$, we have by induction, $\bm{Q}(t) \le  \bm{Q}^*(t)$. 

By a similar argument, given $\mathscr{E}_{\ref{ev:accurate-est}} \cap \mathscr{E}_{\ref{ev:no-explore}} \cap \mathscr{E}_{\ref{ev:cmu-hit-zero}}$, we have $\bm{Q}^*(t) \le  \bm{Q}(t)$. Combining the two, gives us the required result, i.e., $\PP\bigl[\bm{Q}(t) = \bm{Q}^*(t) \,|\,
\mathscr{E}_{\ref{ev:accurate-est}} \cap \mathscr{E}_{\ref{ev:no-explore}}
\cap \mathscr{E}_{\ref{ev:cmuhat-hit-zero}} \cap \mathscr{E}_{\ref{ev:cmu-hit-zero}}\bigr]=1$ a.s.
\Halmos\endproof

We are now ready for the proof of \cref{thm:c-mu-hat}.

\proof{Proof of \cref{thm:c-mu-hat}.}
For any norm function $\norm{\cdot}$, given \cref{lem:correct-order,lem:no-explore,lem:cmuhat-hit-zero,lem:cmu-hit-zero}, we have
\begin{equation*}
\EE\left[ \bnorm{\mathbf{Q}(t) - \mathbf{Q}^*(t)}_1 \right] \le \bigl( \PP\left[ \mathscr{E}_{\ref{ev:accurate-est}}^c \right] + \PP\left[ \mathscr{E}_{\ref{ev:accurate-est}} \cap \mathscr{E}_{\ref{ev:no-explore}}^c \right] + \PP\left[ \mathscr{E}_{\ref{ev:accurate-est}} \cap \mathscr{E}_{\ref{ev:cmuhat-hit-zero}}^c \right] + \PP\left[ \mathscr{E}_{\ref{ev:cmu-hit-zero}}^c \right] \bigr) (\bnorm{\mathbf{Q}(0) + t \mathbf{1}}) = o\left( \frac{1}{t^k} \right),
\end{equation*}
which gives us
\begin{equation*}
\lim_{t \to \infty} t^k \EE\left[ \bnorm{\mathbf{Q}(t) - \mathbf{Q}^*(t)}_1 \right] = 0.
\end{equation*}
The above result also implies that the regret scales as a constant with $T$. This can be seen as follows:
\begin{equation*}
\Psi(T) = \sum_{t=1}^T \EE\left[ \sum_{i=1}^U c_i \bigl( Q_i(t) - Q^*_i(t)\bigr) \right] \le \left( \max_{i\in [U]} c_i \right) \sum_{t=1}^T \EE\left[ \bnorm{\mathbf{Q}(t) - \mathbf{Q}^*(t)}_1 \right] \le C,\qquad\Halmos
\end{equation*}
for sufficiently large $T$ and some constant $C$ which is independent of $T$.
\endproof

\begin{remark}
\label{rem:ext2}
It is easy to see that the above proof holds for discounted cost
under any discount factor.
\end{remark}

\section{Proofs of
\texorpdfstring{\cref{lem:correct-order,lem:no-explore,lem:cmuhat-hit-zero}}{}.}\label{AppD}
\proof{Proof of \cref{lem:correct-order}.}
Let $L \df \lfloor \sqrt{t} \rfloor$ and
$\Gamma(L) \df \max \{l \le L : \varepsilon(l) = 0\}$. We first show that $N_{\mathsf{min}}(L) \ge \log^2 t$.
Note that $\Gamma(L) > 0$ implies that
$N_{\mathsf{min}}(\Gamma(L)) \ge \Upsilon(\Gamma(L))$.

Now, fix some arbitrary sequence to the $U$ assignments in
$\mathscr{E}$ and let $\left\lbrace X(s), s \in \NN  \right\rbrace$
denote a sequence of i.i.d.\ discrete random variables which are
uniformly distributed value in $[U]$.
These random variables denote the assignment chosen when the algorithm decides
to explore. Note that
\begin{equation*}
N_{\mathsf{min}}(L) \ge N_{\mathsf{min}}(\Gamma(L)) + \min_{i \in [U]}
\Biggl\lbrace \sum_{s=\Gamma(L)+1}^L \mathsf{B}(s) \ind{X(s) = i} \Biggr\rbrace.
\end{equation*} 
Therefore, $\left\lbrace N_{\mathsf{min}}(L) < \log^2 t \right\rbrace \cap \left\lbrace \Gamma(L) = l \right\rbrace$ (for some $l \in [L]$) implies that
\begin{equation*}
\min_{i \in [U]} \left\lbrace \sum_{s=l+1}^L \mathsf{B}(s) \ind{X(s) = i} \right\rbrace < \log^2 t - \Upsilon(l).\end{equation*}
This gives us
\begin{align}\label{eq:blah}
\PP\left[ N_{\mathsf{min}}(L) < \log^2 t \right] & = \sum_{l=0}^L \PP\left[ \left\lbrace N_{\mathsf{min}}(L) < \log^2 t \right\rbrace \cap \left\lbrace \Gamma(L) = l \right\rbrace \right]	\nonumber\\[3pt]
& \le \sum_{l=0}^L \sum_{i \in [U]} \PP\left[ \sum_{s=l+1}^L \mathsf{B}(s) \ind{X(s) = i} < \log^2 t - \Upsilon(l) + 1 \right].	
\end{align}
Now, note that for any $i \in [U]$, $\left\lbrace \mathsf{B}(s) \ind{X(s) = i}, s \in \NN  \right\rbrace$ are independent Bernoulli random variables with mean
\begin{equation*}
\EE\left[ \mathsf{B}(s) \ind{X(s) = i} \right] = \min \left\lbrace 1, 3 \frac{\log^2l}{l} \right\rbrace.
\end{equation*}
The sum of their means can be lower bounded as follows. Let \const{const:1}$C_{\ref{const:1}} := \min \left\lbrace l : \frac{1}{3} \ge \frac{\log^2x}{x} \right\rbrace$. For any $t_2 > t_1 > 0$,
\begin{align*}
\EE\left[ \sum_{l = t_1 + 1}^{t_2} \mathsf{B}(l) \ind{X(l) = l} \right] 
& = \sum_{l = t_1 + 1}^{t_2} \min \left\lbrace 1, 3 \frac{\log^2l}{l} \right\rbrace
\nonumber	\\[5pt]
& \ge 3  \int_{t_1 + 1}^{t_2 + 1} \frac{\log^2l}{l} \diff l - C_{\ref{const:1}} \qquad \Bigl(\because 3 \frac{\log^2l}{l} \le 2\ \forall l\Bigr)	\nonumber	\\[5pt]
& = \log^3 (t_2 + 1) - \log^3 (t_1 + 1) - C_{\ref{const:1}}.
\end{align*}
Similarly, we can also compute the following upper bound.
\begin{equation*}
\EE\left[ \sum_{l = t_1 + 1}^{t_2} \mathsf{B}(l) \ind{X(l) = l} \right]  \le
\left(\log^3 t_2 - \log^3 t_1 \right).
\end{equation*}
Now, for $L = \lfloor \sqrt{t} \rfloor$  and for any $l \in \{0, 1, \dotsc, L\}$, let
\begin{equation*}
\xi_l \df \EE\left[ \sum_{s=l+1}^{L} \mathsf{B}(s) \ind{X(s) = l} \right]\,.
\end{equation*}
Using the above bounds, we get
\begin{equation*}
\log^3 (\sqrt{t}) - \log^3 (l + 1) - C_{\ref{const:1}} \, \le \, \xi_l \, \le \, \log^3 (\sqrt{t}) \quad \forall 0 \le l \le L.
\end{equation*}
Let \const{const:2}$C_{\ref{const:2}} := \max \left\lbrace l : \frac{1}{8}\log^3 l < \log^{5/2} l + \log^2 l + C_{\ref{const:1}} + 3\right\rbrace$. Then, for $t > C_{\ref{const:2}}$, we have
\begin{align*}
\xi_l - \sqrt{8 \log^2 t \xi_l }
& \ge \log^3 (\sqrt{t}) - \log^3 (l + 1) - C_{\ref{const:1}} - \log^{5/2} t	\\[5pt]
& \ge \log^2 t - \Upsilon(l) + 1 ,
\end{align*}
where the last inequality follows by the definition of $C_{\ref{const:2}}$ and using the fact that $\log^3 (l + 1) - 2 \log^3 (l-1) \le \log^3 (l + 1) -  \Upsilon(l) \le 2 \; \forall l \ge 2$. Now, using the Chernoff bound for independent Bernoulli variables, we have
\begin{align*}
\PP\left[ \sum_{s=l+1}^L \mathsf{B}(s) \ind{X(s) = i} < \log^2 t - \Upsilon(l) + 1 \right]
& \le \PP\left[ \sum_{s=l+1}^L \mathsf{B}(s) \ind{X(s) = i} \le \xi_l - \sqrt{8 \log^2 t \xi_l } \right]	\\[3pt]
& \le \exp\left( -4\log^2 t \right).
\end{align*}
Using this in \cref{eq:blah}, we get
\begin{equation}\label{eq:samples-lb}
\PP\left[ N_{\mathsf{min}}(L) < \log^2 t \right]   = o\left( \frac{1}{t^{k+1}} \right).
\end{equation}
Let $R_{i,j,n}$ denote the outcome ($1$ if success, $0$ otherwise) of the
$n^{\mathrm{th}}$ assignment of the server $j$ to queue $i$. Note that $\{R_{i,j,n}\}_{n \ge 1}$ are i.i.d.\ Bernoulli random variables with mean $\mu_{i,j}$. Let $\hat{\mu}_{i,j,n}$ be the average number of successes in the first $n$ assignments of the server $j$ to queue $i$, i.e.,
$\hat{\mu}_{i,j,n} = \frac{1}{n}\sum_{i=1}^n R_{i,j,n}$.
Recall the definition of $\varDelta$ in \cref{E-Delta}.
Using Hoeffding's inequality for i.i.d.\ Bernoulli random variables, 
we obtain
\begin{align*}
\sum_{i \in [U], j \in [K]} \sum_{n = \log^2 t}^{\infty}
\PP\bigl[ c_i \card{\hat{\mu}_{i,j,n} - \mu_{i,j}} \ge \nicefrac{\varDelta}{2} \bigr]
& \le \sum_{i \in [U], j \in [K]} \sum_{n = \log^2 t}^{\infty} 2 \exp \left( -\frac{\varDelta^2}{2} n  \right)	\\[3pt]
 & \le 2 U K  \exp \left( -\frac{\varDelta^2}{2} \log^2 t  \right) \sum_{n = 0}^{\infty} \exp \left( -\frac{\varDelta^2}{2} n  \right)		\\[3pt]
 & = \frac{2 U K \exp \left( -\frac{\varDelta^2}{2} \log^2 t  \right)}{1-\exp\left(  -\frac{\varDelta^2}{2} \right)}	\\
 & = o\left( \frac{1}{t^{k+1}} \right).
\end{align*}
Combining the above inequality with \cref{eq:samples-lb}, we have, 
\begin{equation*}
\PP\left[ \mathscr{E}_{\ref{ev:accurate-est}}^c \right]  \le \PP\left[ N_{\mathsf{min}}(L) < \log^2 t \right] + \sum_{i \in [U], j \in [K]} \sum_{n = \log^2 t}^{\infty}
\PP\bigl[ c_i \card{\hat{\mu}_{i,j,n} - \mu_{i,j}} \ge
\nicefrac{\varDelta}{2} \bigr] = o\left( \frac{1}{t^{k+1}} \right).\qquad\Halmos
\end{equation*}
\endproof

\subsection{A coupled queueing system.}
 In order to prove \cref{lem:no-explore,lem:cmuhat-hit-zero}, we construct a coupled queueing system which has $U$ queues and $K$ servers with the same link rate distribution as the original system---given by $\mathrm{Bernoulli}(\bm{\mu})$.
Denote the queue-length of the coupled system at any time $l$ by
$\widetilde{\bm{Q}}(l)$.
For $1 \le l \le \lfloor \sqrt{t} \rfloor + 1$, we have
$\widetilde{\bm{Q}}(l) = \bm{Q}(1) + (l-1) \bm{1}$.
For $l > \sqrt{t}$, the system evolves as follows.
Recall that $\{\mathsf{B}(l),\, l = 1, 2 \dotsc\}$ are independent Bernoulli
samples used in the $c\hat{\mu}$ algorithm to choose between exploration and
exploitation. Arrivals to the coupled queueing system are given by
\begin{equation*}
\tilde{\bm{A}}(l) = \mathsf{B}(l) \oplus \bm{A}(l)
\quad \forall l > \sqrt{t},
\end{equation*}
where $\oplus$ denotes the binary XOR operator.
In every time-slot $l > \sqrt{t}$, servers are assigned to the queues in the
coupled system according to the $c\mu$ rule.
Therefore, starting from time $\lfloor \sqrt{t} \rfloor + 1$,
the process $\{\widetilde{\bm{Q}}(l)\}_{l > \sqrt{t}}$ represents the
queue-length evolution of a $c\mu$ system with arrival rates
$\{\widetilde{\bm{Q}}(l)\}_{l > \sqrt{t}}$.

Recall that the service offered by each of the links is i.i.d.\ across time, with mean given by the rate matrix $\bm{\mu}$.
We use the alternate construction of the service process for the original and coupled systems used in the proof of \cref{lem:actual-genie-monotonic}.
This results in the same marginal distribution for both the systems as the i.i.d.\ service process.
In analogy to \cref{lem:actual-genie-monotonic}, we can show that
for this modification of the service process the queue-lengths given by the
$c\hat{\mu}$ algorithm do not exceed the queue-lengths in the coupled queueing system.
This is stated in the following lemma, whose proof is very similar to that of \cref{lem:actual-genie-monotonic}.

\begin{lemma}\label{lem:actual-coupled-monotonic}
We have $\PP\bigl[\bm{Q}(l) \le \widetilde{\bm{Q}}(l)\,|\,
\mathscr{E}_{\ref{ev:accurate-est}}\bigr]=1$ for all $1 \leq l \leq t$.
\end{lemma}

We use \cref{lem:actual-coupled-monotonic} in the proofs
of \cref{lem:no-explore,lem:cmuhat-hit-zero} to obtain tail bounds on the busy cycles
of the $c\hat{\mu}$ system.
Let
\begin{equation}\label{E-beta}
\theta \df \min_{i \in [U]} \left( \lambda_i \Pi_{i' \in [U], i' \neq i}
\left( 1 - \lambda_{i'} \right) \right)^K \left( \Pi_{j \in [K]}
\left( 1 - \mu_{i,j} \right) \right)^{K-1}.
\end{equation}

\proof{Proof of \cref{lem:no-explore}.}
We prove this lemma in two steps:
\begin{enumerate}[label=(\alph*)]
\item \label{item:num-busy-cycle} 
Let
\begin{equation*}
\mathscr{E}_{\ref{ev:no-explore}\ref{item:num-busy-cycle}} \df
\Biggl\lbrace \sum_{l=1}^{\nicefrac{t}{2}} \ind{\bm{Q}(l) = \bm{0}}
\ge \lfloor \sqrt{t} \rfloor \Biggr\rbrace.\end{equation*}
Then $\PP\bigl[ \mathscr{E}_{\ref{ev:no-explore}\ref{item:num-busy-cycle}}^c \cap \mathscr{E}_{\ref{ev:accurate-est}}  \bigr] = o\left( \frac{1}{t^{k+1}} \right)$.
\item \label{item:samples-lb2} For any $l \in \NN$, $i \in [U]$, let $\mathscr{E}^{l,i}$ be the event that, in the first $K$ slots of the $l$th busy cycle, there are no arrivals to any queue except queue $i$ which has $K$ arrivals, and in the first $K-1$ slots, all servers have zero service to queue $i$. Define the Bernoulli random variables 
$X_{l,i} \df \ind{\mathscr{E}^{l,i}}$,
and the event
\begin{equation*}\mathscr{E}_{\ref{ev:no-explore}\ref{item:samples-lb2}} \df
\Biggl\lbrace \min_{i \in [U]} \sum_{l=1}^{\lfloor \sqrt{t} \rfloor} X_{l,i} \ge \frac{\theta}{2}\lfloor \sqrt{t} \rfloor \Biggr\rbrace,\end{equation*}
where $\theta$ is as in \cref{E-beta}.
Then $\PP\bigl[ \mathscr{E}_{\ref{ev:no-explore}\ref{item:samples-lb2}}^c \bigr] = o\left( \frac{1}{t^{k+1}} \right)$.
\end{enumerate}
Consider any $t$ such that
\begin{equation*}
\lfloor \sqrt{t} \rfloor  \ge \max \left\lbrace 2\left( C_{\ref{const:3}} (k+2) \log t + C_{\ref{const:4}}  + C_{\ref{const:5}} + 1 \right), \frac{4}{\theta} \log^3 (t) \right\rbrace\,.
\end{equation*}
That $\mathscr{E}_{\ref{ev:no-explore}\ref{item:num-busy-cycle}} \cap \mathscr{E}_{\ref{ev:no-explore}\ref{item:samples-lb2}}$ implies $\mathscr{E}_{\ref{ev:no-explore}}$ can be seen as follows : since  $\lfloor \sqrt{t} \rfloor \ge \frac{4}{\theta} \log^3 (t)$, given $\mathscr{E}_{\ref{ev:no-explore}\ref{item:num-busy-cycle}} \cap \mathscr{E}_{\ref{ev:no-explore}\ref{item:samples-lb2}}$, for any $l \in (\nicefrac{t}{2}, t]$, we have
 \begin{equation*}N_{\mathsf{min}}(l) \ge N_{\mathsf{min}}(\nicefrac{t}{2}+1) \ge \frac{\theta}{2}\lfloor \sqrt{t} \rfloor \ge \Upsilon(l).\end{equation*} This implies that $\varepsilon(l) = 0$  $\forall l \in (\nicefrac{t}{2}, t]$.
 Therefore,
 \begin{equation*}\PP\left[ \mathscr{E}_{\ref{ev:no-explore}}^c  \cap \mathscr{E}_{\ref{ev:accurate-est}} \right] \le \PP\left[ \mathscr{E}_{\ref{ev:no-explore}\ref{item:num-busy-cycle}}^c  \cap \mathscr{E}_{\ref{ev:accurate-est}} \right] + \PP\left[ \mathscr{E}_{\ref{ev:no-explore}\ref{item:samples-lb2}}^c \right] = o\left( \frac{1}{t^{k+1}} \right).\end{equation*}
 
To prove part~\ref{item:num-busy-cycle}, note that, by
\cref{lem:actual-coupled-monotonic}, we have
 \begin{equation*}
 \PP\left[ \mathscr{E}_{\ref{ev:no-explore}\ref{item:num-busy-cycle}}^c \cap \mathscr{E}_{\ref{ev:accurate-est}}  \right] \le \PP\Biggl[ \sum_{l=1}^{\nicefrac{t}{2}} \ind{\widetilde{\bm{Q}}(l) = \bm{0}} < \lfloor \sqrt{t} \rfloor \Biggr].
\end{equation*} 
Now, let $\tilde{\tau}_1, \tilde{\tau}_2, \dotsc, \tilde{\tau}_{\lfloor \sqrt{t} \rfloor}$ be the first $\lfloor \sqrt{t} \rfloor$ busy cycle lengths of the coupled queueing system. Since the coupled queueing system behaves like the $c\mu$ system after time $\sqrt{t}$, we can use \cref{lem:conc-busy-period} to obtain tail bounds for its busy periods. Using that $\left( C_{\ref{const:3}} (k+2) \log t + C_{\ref{const:4}}  + C_{\ref{const:5}} + 1 \right) \sqrt{t} \le \nicefrac{t}{2}$, we have
\begin{align*}
& \PP\Biggl[ \sum_{l=1}^{\nicefrac{t}{2}} \ind{\widetilde{\bm{Q}}(l) = \bm{0}} < \sqrt{t} \Biggr]	\\
& \le \PP\left[ \tilde{\tau}_1 >  \sqrt{t} + C_{\ref{const:3}} (k+2) \log t
+  C_{\ref{const:4}} 
\sqrt{t} + C_{\ref{const:5}} \right]
+ \sum_{m=2}^{\lfloor \sqrt{t} \rfloor} \PP\left[ \tilde{\tau}_m > C_{\ref{const:3}} (k+2) \log t + C_{\ref{const:5}} \right]	\\
& \le \frac{1}{t^{k+2}} + \frac{\sqrt{t}}{t^{k+2}}
 = o\left( \frac{1}{t^{k+1}} \right).
\end{align*}

 We now prove part~\ref{item:samples-lb2}. 
 Note that for any given $i$, $X_{l,i}, \; l \in \NN$ are i.i.d.\ Bernoulli variables with mean
 \begin{equation*}
 \EE\left[ X_{l,i} \right] = \left( \lambda_i \Pi_{i' \in [U], i' \neq i} \left( 1 - \lambda_{i'} \right) \right)^K \left( \Pi_{j \in [K]} \left( 1 - \mu_{i,j} \right) \right)^{K-1}.
 \end{equation*}
 Therefore, using part~\ref{item:num-busy-cycle} and the Chernoff bound for the sum of these Bernoulli variables, we obtain
 \begin{align*}
 \PP\left[ \mathscr{E}_{\ref{ev:no-explore}\ref{item:samples-lb2}}^c \right] \le \sum_{i \in [U]} \PP\Biggl[ \sum_{l = 1}^{\lfloor \sqrt{t} \rfloor} X_{l,i} < \frac{\theta}{2}\lfloor \sqrt{t} \rfloor  \Biggr] \le U \exp\left( \frac{-\theta}{8} \sqrt{t} \right) = o\left( \frac{1}{t^{k+1}} \right).
\end{align*}  
This proves the lemma.
\Halmos\endproof

\proof{Proof of \cref{lem:cmuhat-hit-zero}.}
Using \cref{lem:actual-coupled-monotonic}, we have
\begin{equation*}
\PP\left[ \mathscr{E}_{\ref{ev:cmuhat-hit-zero}}^c \cap \mathscr{E}_{\ref{ev:accurate-est}}\right]  \le \PP\left[ \widetilde{\bm{Q}}(l) \neq \bm{0} \quad\forall l \in (\nicefrac{t}{2}, t]\right].
\end{equation*}
Consider any $t$ such that
$\lfloor \sqrt{t} \rfloor \ge 2\left( C_{\ref{const:3}} (k+3) \log t + C_{\ref{const:4}}  + C_{\ref{const:5}} + 1 \right)$, and let $\tilde{\tau}_1, \tilde{\tau}_2, \dotsc, \tilde{\tau}_{\nicefrac{t}{2}+1}$ be the first $\nicefrac{t}{2}+1$ busy cycle lengths of the coupled queueing system. Then, since $C_{\ref{const:3}} (k+3) \log t + \left( C_{\ref{const:4}}  + 1 \right) \sqrt{t} + C_{\ref{const:5}} \le \nicefrac{t}{2}$, the event $\left\lbrace \tilde{\tau}_1 \le  \sqrt{t} + C_{\ref{const:3}} (k+3) \log t + C_{\ref{const:4}}  \sqrt{t} + C_{\ref{const:5}} \right\rbrace$ implies that the first busy period ends before time $\nicefrac{t}{2}$ and the event $\cap_{m=2}^{\nicefrac{t}{2}+1} \left\lbrace \tilde{\tau}_m \le C_{\ref{const:3}} (k+3) \log t + C_{\ref{const:5}} \right\rbrace$ implies that $\widetilde{\bm{Q}}(l)$ hits the zero state at least once in $(\nicefrac{t}{2}, t]$ after the first busy cycle.

By \cref{lem:conc-busy-period}, we have
$\PP\left[ \tilde{\tau}_1 > C_{\ref{const:3}} (k+3) \log t + \left( C_{\ref{const:4}}  + 1 \right) \sqrt{t} + C_{\ref{const:5}} \right] \le \nicefrac{1}{t^{k+3}}$, and similarly, for all $m > 1$, $\PP\left[ \tilde{\tau}_m > C_{\ref{const:3}} (k+3) \log t + C_{\ref{const:5}} \right] \le \nicefrac{1}{t^{k+3}}$. The union bound gives us
\begin{align*}
\PP\left[ \widetilde{\bm{Q}}(l) \neq \bm{0} \; \forall l \in (\nicefrac{t}{2}, t]\right] & \le  \PP\left[ \tilde{\tau}_1 > C_{\ref{const:3}} (k+3) \log t + \left( C_{\ref{const:4}}  + 1 \right) \sqrt{t} + C_{\ref{const:5}} \right]\\
&\mspace{100mu}
+ \sum_{m=2}^{\nicefrac{t}{2}+1} \PP\left[ \tilde{\tau}_m > C_{\ref{const:3}} (k+3) \log t + C_{\ref{const:5}} \right]
 = o\left( \frac{1}{t^{k+1}} \right).\qquad\Halmos
\end{align*}
\endproof

\section{Stationary Distribution of a Single-Queue Two-Server System.}
\label{sec:dist-1x2}
In this section, we derive expressions for the stationary distribution of the queue process in a $2 \times 1$ system. Let $\lambda$ be the arrival rate and $\mu_1, \mu_2$ be the server rates such that $\mu_1$ is given higher priority. The transition probabilities (for states higher than $0, 1$) are
\begin{align*} 
p_{+1} & = \lambda (1-\mu_1) (1-\mu_2);	\\
p_{-1} & = (1-\lambda)( (1-\mu_1)\mu_2 + (1-\mu_2)\mu_1) + \lambda \mu_1 \mu_2;	\\
p_{-2} & = (1-\lambda) \mu_1 \mu_2.
\end{align*}
The balance equations for the Markov chain are given by:
\begin{align*}
\lambda \uppi_0 & = (1-\lambda)\mu_1 \uppi_1 + p_{-2} \uppi_2;	\\
\lambda (1-\mu_1) \uppi_1 & = (p_{-1} + p_{-2}) \uppi_2 + p_{-2} \uppi_3;	\\	
p_{+1} \uppi_i & = (p_{-1} + p_{-2}) \uppi_{i+1} + p_{-2} \uppi_{i+2}
\quad \forall i \ge 2.
\end{align*}
If we can find a root $\alpha \in (0,1)$ for the quadratic equation
\begin{equation*}
p_{+1}  = (p_{-1} + p_{-2}) x + p_{-2} x^2,
\end{equation*}
then we can get a closed form expression for $\bm{\uppi}$ which satisfies
\begin{align*}
\uppi_{i+1} & = \alpha \uppi_{i} \quad \forall i \ge 2;	\\
\lambda (1-\mu_1) \uppi_1 & = \left( (p_{-1} + p_{-2}) + p_{-2} \alpha \right) \uppi_{2};	\\
\lambda \uppi_0 & = (1-\lambda)\mu_1 \uppi_1 + p_{-2} \uppi_2;	\\
\bm{\uppi}^T \bm{1} & = 1.
\end{align*}
For this, we need
\begin{equation} \label{eq:root}
\begin{aligned}
\alpha \,\df\, \frac{1}{2 p_{-2}} &\left(  -(p_{-1} + p_{-2}) + \sqrt{ (p_{-1} + p_{-2})^2 + 4 p_{+1} p_{-2}} \right) < 1			\\
\iff & \sqrt{ (p_{-1} + p_{-2})^2 + 4 p_{+1} p_{-2}} < (p_{-1} + p_{-2}) + 2 p_{-2}	\\
\iff & (p_{-1} + p_{-2})^2 + 4 p_{+1} p_{-2} < (p_{-1} + p_{-2})^2 + 4 (p_{-1} +  p_{-2}) p_{-2} + 4 p_{-2}^2	\\
\iff & p_{+1} < p_{-1} + 2p_{-2}	\\
\iff & \lambda < \mu_1 + \mu_2.
\end{aligned}
\end{equation}
Therefore, for any stable system, for $\alpha$ given in \cref{eq:root}, we have 
$\uppi_{i+1}  = \alpha \uppi_{i}$ for all $i\ge2$,
\begin{equation*}
\frac{p_{+1}}{\alpha} \uppi_2  = \lambda (1-\mu_1) \uppi_1,\quad
\left( (1-\lambda)\mu_1 + \frac{\alpha p_{-2}}{1-\mu_2} \right) \uppi_1
= \lambda \uppi_0,\quad\text{and\ \ }
\uppi_0 + \uppi_1 + \frac{\uppi_2}{1-\alpha} = 1,
\end{equation*}
from which we obtain
\begin{align*}
\uppi_0 & = \left( 1 + \left( \frac{\lambda}{(1-\lambda)\mu_1 + \frac{\alpha p_{-2}}{1-\mu_2}} \right)\left( 1 + \frac{\alpha}{(1-\alpha)(1-\mu_2)} \right) \right)^{-1};
\\[5pt]
\uppi_0 + \uppi_1 & = \left( 1 + \frac{\lambda}{(1-\lambda)\mu_1 + \frac{\alpha p_{-2}}{1-\mu_2}} \right) \uppi_0.	
\end{align*}
\end{APPENDICES}

\section*{Acknowledgments.}
This work was partially supported by NSF grants CNS-1343383 and
DMS-1715210, Army Research Office grant W911NF-17-1-0359 and
W911NF-17-1-0019, Office of Naval Research grant N00014-16-1-2956, and
the US DoT supported D-STOP Tier 1 University Transportation Center.

\bibliographystyle{informs2014}
\bibliography{./Learn-cmu}
\end{document}